\documentclass[manuscript,screen]{acmart}

\AtBeginDocument{%
  \providecommand\BibTeX{{%
    \normalfont B\kern-0.5em{\scshape i\kern-0.25em b}\kern-0.8em\TeX}}}

\setcopyright{acmcopyright}

\copyrightyear{2021}

\acmConference[TOSEM]{TOSEM: ACM Transaction on Software Engineering and Methodology}{Jan 20--21, 2022}{Woodstock, NY}
\acmBooktitle{Woodstock '18: ACM Symposium on Neural Gaze Detection,
  June 03--05, 2018, Woodstock, NY}
\acmPrice{15.00}
\acmISBN{978-1-4503-XXXX-X/18/06}

\usepackage{svg}
\usepackage{amsmath}  
\usepackage{xcolor}   
\usepackage{tabularx}
\newcommand{\specialcell}[2][l]{%
  \begin{tabular}[#1]{@{}l@{}}#2\end{tabular}}
\usepackage[flushleft]{threeparttable} 
\usepackage{booktabs,caption}
\usepackage{graphicx}
\newcolumntype{C}[1]{>{\centering}p{#1}}
\usepackage{microtype}
\usepackage{listings}

\usepackage{mathtools}
\usepackage{blkarray, bigstrut}
\usepackage{gauss}

\usepackage{multirow}
\usepackage{multicol}
\usepackage{longtable}

\begin{document}

\title{APIRO: A Framework for Automated Security Tools API Recommendation}

\author{Zarrin Tasnim Sworna}
\affiliation{%
  \institution{CREST – Centre for Research on Engineering Software Technologies, University of Adelaide and Cyber Security Cooperative Research Centre (CSCRC)}
 \country{Australia}}
  \email{zarrintasnim.sworna@adelaide.edu.au}

\author{Chadni Islam}
\affiliation{%
  \institution{CREST–Centre for Research on Engineering Software Technologies, University of Adelaide}
 \country{Australia}
} \email{chadni.islam@adelaide.edu.au}

\author{Muhammad Ali Babar}
\affiliation{%
  \institution{CREST – Centre for Research on Engineering Software Technologies, University of Adelaide and Cyber Security Cooperative Research Centre (CSCRC)}
 \country{Australia}
} \email{ali.babar@adelaide.edu.au}


\begin{abstract}
Security Orchestration, Automation, and Response (SOAR) platforms integrate and orchestrate a wide variety of security tools to accelerate the operational activities of Security Operation Center (SOC). Integration of security tools in a SOAR platform is mostly done manually using APIs, plugins, and scripts. SOC teams need to navigate through API calls of different security tools to find a suitable API to define or update an incident response action. Analyzing various types of API documentation with diverse API format and presentation structure involves significant challenges such as data availability, data heterogeneity, and semantic variation for automatic identification of security tool APIs specific to a particular task. Given these challenges can have negative impact on SOC team's ability to handle security incident effectively and efficiently, we consider it important to devise suitable automated support solutions to address these challenges. We propose a novel learning-based framework for automated security tool \underline{\textbf{API}} \underline{\textbf{R}}ecommendation for security \underline{\textbf{O}}rchestration, automation, and response, \textbf{APIRO}. To mitigate data availability constraint, APIRO enriches security tool API description by applying a wide variety of data augmentation techniques. To learn data heterogeneity of the security tools and semantic variation in API descriptions, APIRO consists of an API-specific word embedding model and a Convolutional Neural Network (CNN) model that are used for prediction of top 3 relevant APIs for a task. We experimentally demonstrate the effectiveness of APIRO in recommending APIs for different tasks using 3 security tools and 36 augmentation techniques. Our experimental results demonstrate the feasibility of APIRO for achieving 91.9\% Top-1 Accuracy. Compared to the state-of-the-art baseline, APIRO is 26.93\%, 23.03\%, and 20.87\% improved in terms of Top-1, Top-2, and Top-3 Accuracy and outperforms the baseline by 23.7\% in terms of Mean Reciprocal Rank (MRR).

\end{abstract}




\maketitle

\section{Introduction}
Security Operation Centers (SOCs) are increasingly adopting Security Orchestration, Automation, and Response (SOAR) platforms to manage their operational activities in an integrated fashion. A SOC typically utilizes a combination of different security tools to detect, analyze, and respond to cyber security incidents \cite{esg}. A SOAR platform orchestrates and automates the activities of these security tools such as download malware sample, scan network packets, and kill malicious process. Security tools come with numerous APIs, plugins, and commands that are used for integrating them into a SOAR platform \cite{ Splunk_Phantom, threatconnect_playbook,  sec_orch_caise}. For example, Malware Information Sharing Platform, MISP \cite{PyMISP_apidoc, misp_automation_api, PyMISP_lib} and Endpoint Detection and Response (EDR) tool, Limacharlie \cite{lima_rest, lima_sensor_com, lima_python} have around 398 and 272 APIs, respectively. 
SOAR platforms heavily rely on security tools' APIs to drive orchestration and automation of response actions \cite{Splunk_Phantom, sec_orch_caise, orange_cyber}. For instance, Phantom's SOAR platform uses more than 1900 APIs to execute security tools' activities based on different Incident Response Plans (IRP)\footnote{IRP is the sequence of activities to be executed by security team and security tools to respond to a specific security incident.} \cite{Splunk_Phantom}. Table \ref{tab:irp_api} shows an example IRP of Rapid7 SOAR platform for threat hunting \cite{rapid7}. It shows the tasks and the relevant APIs of multiple security tools that are needed to execute these IRP activities, where the execution of an activity in an IRP may require multiple tools. For example, the activity `\textit{investigation of datasets (Endpoint, Network)}' requires API from Limacharlie and Snort (i.e., a Network Intrusion Prevention System (NIPS)) to scan organizations' data and network packets. To define, update, or execute IRP activities SOC team must sufficiently understand the corresponding API of the relevant security tools \cite{sec_orch_caise}. 
 
Information about security tools' APIs are mostly acquired from security tools' manual, documentation, or websites, which is tedious and time-consuming causing delayed response to security incidents \cite{apibot}. To find a desired piece of API information, security teams may need to read through numerous documents, which is not preferred by security teams \cite{doc_read1, doc_read2}. For example, a SOC team using 25 security tools may require to search through 25 different websites/API manual for finding the information required for a set of suitable APIs when selecting and integrating the relevant tools to deal with a security incident. To support the task of finding suitable APIs, online Question Answering (Q\&A) forums like Stack Overflow (SO), Q\&A forums by security tools' vendors have become quite popular, where one can directly ask a question to the relevant community. However, the responses to these questions are not instant and many are unanswered. For example, in Q\&A forum of Security Information and Event Management (SIEM) tool, Splunk, a user had to wait from 1.5 hours to 16 hours\footnote{https://community.splunk.com/t5/Getting-Data-In/How-to-execute-a-saved-and-on-demand-search-using-REST-API/m-p/29306} to get a response which often refers to the corresponding API documentation. Moreover, around 1225 questions on Splunk Q\&A forum are unanswered\footnote{https://community.splunk.com/t5/Using-Splunk/ct-p/use-splunk}. 
Hence, there is a critical need for an automatic recommendation system for security team to instantly find API related information to support the selection and integration of different tools for timely responses to security incidents.

The existing studies \cite{from_word_to_doc, biker, biker_tool, rack, mulapi} on query-based API recommendation extensively utilize Natural Language Processing (NLP) techniques such as similarity measure, word embedding, and IDF. They mainly recommend APIs for a specific language (e.g., Java). Word embedding and Inverse Document Frequency (W2V-IDF) based similarity measure have been adopted by Ye et al. \cite{from_word_to_doc} and Huang et al. \cite{biker} for Java API recommendation.
Since the same concept can be presented using different vocabularies in a query, to mitigate this vocabulary mismatch problem Rahman et al. \cite{rack} have used SO Q\&A forum data related to Java APIs. Huang et al. \cite{biker} have used both SO and API documentation data to develop their proposed recommendation approach. Though these techniques perform well for language specific problems, there are challenges related to data heterogeneity, data availability, and semantic variation to apply them for recommending security tools API.

Firstly, the existing approaches on API recommendation \cite{rack, biker, portfolio} are language specific and considered similar types of APIs (e.g., Java or C). They do not consider building a unified model for different programming languages. However, different security tools have different types of documentation for APIs, that varies in terms of type of API (e.g., REST API, Python API class, Python API method, and commands), data source (e.g., PDF, HTML format), and documentation structure (e.g., swagger API documentation structure, descriptive documentation structure). The existing approaches (\cite{from_word_to_doc, biker}) have built language specific word embedding model such as Java specific model. Following this approach require SOC to build multiple tool-specific word embedding model for each security tool which is not feasible. Hence, a unified framework is required to answer queries regarding any security tool for different APIs.

Secondly, SO-based approaches can only recommend APIs, that have been highly discussed on SO \cite{biker, rack}. SO does not have similar questions for all APIs of different security tools and many questions on SO remain unanswered. One reason behind less security tool API related questions on SO can be the rapid development of new security tools with new functionalities and APIs. Parnin et al. \cite{SO_less_api} showed that SO is slow to cover new APIs and may ignore important information of an API. Another reason for SOC team not asking questions on SO can be less reliability of SO data, which can lead to vulnerabilities or security flaws in code, whereas API documentation is considered more reliable data-source \cite{SO_insecure}.

Thirdly, the existing approaches (\cite{from_word_to_doc, biker}) have used W2V-IDF based similarity measure, which requires query to have similar keywords to API description in documentation. However, free-form natural language query may have synonyms, paraphrases, semantic and contextual similar words, spelling mistake, and incorrect order of words. Besides, API descriptions in security tool API documentation are too short. For example, mean word count in API description of Limacharlie is 8 and MISP is 13. Also, gathering API description from documentation does not provide semantic variation like SO data. Hence, the similarity score based approach to answer natural language query in such short text with poor semantic variation suffers from low accuracy and vocabulary mismatch problem \cite{slr, mulapi}.


To overcome the above-mentioned three challenges, we propose a framework of \underline{\textbf{API}} \underline{\textbf{R}}ecommendation for security \underline{\textbf{O}}rchestration, automation, and response, namely \textbf{APIRO} to automatically recommend API of multi-vendor, heterogeneous security tools. To overcome data heterogeneity related issues, we have built a unified corpus by developing different adapters. Each adapter use diverse data extraction methods to collect API information from API documentation of varied security tools. Next, to mitigate the data availability constraint, we propose to enrich semantic variation in the security tool API corpus and build an augmented corpus by using a wide variety of data augmentation techniques. 

Similarity score-based (W2V-IDF) approach performs less efficiently for short API description, so we propose to build a Convolutional Neural Network (CNN)-based approach on top of our security tool API-specific word embedding to learn the semantic variation of each API data. To capture the semantics of words (have similar word embeddings for semantically similar words), a security tool API-specific word embedding model covering multiple security tools' API based on the augmented corpus is created. The convolutional layer of the CNN model automatically captures the context and extracts useful features from word embedding model of the API data, which helps efficiently recommending the relevant APIs. Lastly, given a query, APIRO is designed to provide the relevant ranked list of API with corresponding information (i.e., API, description, parameter, and return data which are available in the API documentation). 


We have experimentally evaluated APIRO using API documentation of three different types of security tools (Limacharlie, MISP, and Snort). To experimentally evaluate APIRO, we have devised three Research Questions (RQs) that would help a reader to understand the context and nature of evaluation performed: \textbf{RQ1:} Are data augmentation techniques suitable for API recommendation for natural language query? \textbf{RQ2:} How does APIRO perform compared to the baseline method for security tool API recommendation? \textbf{RQ3:} How effective is APIRO for answering free-form natural language query? We have applied 36 data augmentation techniques and have evaluated the applicability of these augmentation techniques to mitigate data availability constraints. We have considered similarity score based method W2V-IDF \cite{from_word_to_doc, biker, biker_tool} as our baseline approach and have applied it on security tool API data. Experimental results show that APIRO and W2V-IDF can recommend the correct API in Top-1 recommended API for 91.9\% and 72.4\% of the queries, respectively. APIRO outperformed the baseline approaches by 26\%, 23\%, and 20.9\% when considering different Top-N ranking such as Top-1, Top-2, and Top-3 ranked APIs. Hence, APIRO can identify APIs more accurately for different queries compared to the baseline. Getting accurate API recommendation is important to avoid inconsistent and delayed security incident response caused by the use of incorrect APIs. Besides, APIRO achieves Mean Reciprocal Rank (MRR) of 0.94, which outperforms W2V-IDF by 23.7\%. This indicates that APIRO provides the accurate API at a higher rank of the recommended API list. The analysis of the augmented data further shows that data augmentation approaches which preserved the semantics are more suitable for enriching the security tools' API data. 




The main contributions of the research reported in this paper are as follows:
\vspace{-5pt}
\begin{itemize} 
	\item Analysis of the applicability of a wide variety of data augmentation techniques to mitigate the data availability constraint based on an augmented API corpus.
	\item A novel deep neural network (CNN-based) approach to learn the semantic variation of the augmented API corpus that is built using API-specific word embedding and recommend a ranked list of APIs of heterogeneous security tools in response to SOC teams' free-form natural language query.
	\item A quantitative evaluation to evaluate the effectiveness of APIRO using 36 augmentation techniques and 815 APIs gathered from three security tools APIs that gives an augmented corpus with 24,420 diverse API descriptions.
\end{itemize}
\vspace{-5pt}

The rest of the paper is organized as follows: section \ref{sec:motivationScenario} presents a motivation scenario and the related works are presented in Section \ref{sec: related}. Section \ref{sec: framework} presents our proposed framework, APIRO for automated security tool API recommendation. The experimental settings and implementation details are presented in Section \ref{sec: evaluation}. Section \ref{subsec:exp_res} and Section \ref{sec: threats} present the evaluation results and discussion, respectively. Finally, the paper is concluded in Section \ref{sec: concl}.

\begin{table}[h]
\centering
\caption{Example IRP for threat hunting showing the list of activities with APIs of security tools that are required to execute the IRP}
\label{tab:irp_api}
\small
\begin{tabularx}{\textwidth}{p{2.9cm} p{4.3cm} p{1.9cm} p{4.5cm}}
\hline
\textbf{IRP activities } & \textbf{Example tasks} & \textbf{Security tools} & \textbf{Security tools API }\\ \hline

Insert Indicator(s) of  & Download all malware samples & MISP & https:///attributes/downloadSample/
\\ compromise (IOC ) & & & {[}hash{]}/{[}allSamples{]}/{[}eventID{]} \\ \hline
Investigate on datasets & Scan entire organization data & Limacharlie & limacharlie.Replay.Replay.scanEntireOrg() \\ 
(Endpoint, Network)  & Log and scan the network packets and produce alerts & Snort & ./snort -b -A fast -c snort.conf \\ \hline
Generate artifacts to reveal  attacker TTPs & Get malicious file information, timestamps, sizes, etc. & Limacharlie & file\_info {[}-h{]} file \\  
\hline
{Remediate} & Delete malicious file from endpoint. & Limacharlie & file\_del {[}-h{]} file \\ 
 & Drop the malicious network packets & Snort & Config enable\_ipopt\_drops \\ \hline
Enhance detection capabilities & Add a False Positive rule to the Organization. & Limacharlie & limacharlie.Manager.Manager.add\_fp() \\ \hline
\end{tabularx}
\vspace{-20pt}
\end{table}
\section{Motivating scenario} \label{sec:motivationScenario}
Consider the IRP of Rapid7 SOAR platform as shown in Table 1, which presents the sequence of activities and tasks that need to be performed in response to threat hunting. Table 1 shows some security tools, such as Limacharlie, MISP, and Snort, and their corresponding APIs that need to be used for performing the required tasks for a particular type of security incident, e.g., detecting a malicious file on an endpoint. For using the suitable security tools, their relevant APIs need to be understood and linked to 
the tasks. 
Identifying the required API for each task helps to define, update, or modify the IRPs when required. For instance, to define the corresponding tasks for the IRP activity of ‘remediate’, that is ‘delete the malicious file from endpoints’ and ‘drop malicious network packets’, the required tools need to be identified first. After selecting the required tools, e.g., Limacharlie and Snort, it is required to know ‘How to remove the malicious file from endpoint?’ to map the Limacharlie API that deletes a file from the endpoint. Then, for the same ‘remediation’ activity, it is required to know ‘How to drop malicious packets?’ using Snort. Since manually finding suitable APIs from different security tools' documentation is challenging as previously mentioned \cite{doc_read1, doc_read2}, 
there is a need of automated support for analyzing APIs documentation for identifying and recommending suitable APIs that can help to define, update, or execute IRP for the new incidents, new tools, or new APIs of an existing tool. In these types of scenarios, our proposed approach, APIRO, can automatically identify and recommend suitable APIs of the relevant security tools for the required tasks to be performed for an IRP. 

\section{Related Work} \label{sec: related}
A wide spectrum of the related work in the literature addresses tool integration and incident response in SOAR as well as API recommendation for specific programming language. Here we summarize the relevant studies on SOAR, API recommendation, and application of the advanced NLP and DL techniques in API recommendation. 

\subsection{Security Orchestration, Automation, and Response}
SOAR platform aims at integrating multivendor security tools so that a SOC team member can automate the repetitive incident response activities \cite{sec_orch_caise, swimlane, islam2019ontology}. Islam et al. \cite{sec_orch_caise, islam2019ontology} present a semantic based approach to integrate security tools with a SOAR platform. This work considers extracting the required APIs manually from API documentation to create an ontological knowledge base. A SOAR vendor named Swimlane \cite{swimlane} uses API-first architecture for their SOAR platform to support incident response automation, where developers/users of the SOAR platform need to search for the required APIs from documentations. Similarly, two other SOAR vendors D3 \cite{d3_sec} and ThreatConnect \cite{threat_connect} provide SOAR platforms, which leverage APIs for easy integration of tools but they do not provide any support to security teams for finding relevant APIs from diverse security tools. Islam et al. \cite{islam2020architecture} propose architectural support for integrating security tools, requiring manual API learning from documentation. Though the existing works have focused on automation and orchestration of security tool integration and their activities, these studies lack providing suitable support for automatically extracting APIs of security tools. Hence, in this work we propose a unified security tool API recommendation framework to support automated API extraction from API documentations of diverse security tools using natural language query.
\vspace{-8pt}
\subsection{API Recommendation}
The existing research on API recommendation is focused on recommending API for a specific programming language such as Java or C \cite{from_word_to_doc, biker, portfolio} using different approaches (e.g., keyword-API mapping, similarity measure). Rahman et al. \cite{rack} propose RACK, that maps keywords and API for Java API from SO questions and accepts answers. Ye et al. \cite{from_word_to_doc} apply word embedding and IDF-based similarity measure (W2V-IDF) for recommendation of API from API documents and tutorials. Huang et al. \cite{biker} also adopt this same approach (W2V-IDF) for recommending Java API to a query using SO knowledge and API documentation and outperform RACK \cite{rack}. Recently, they have also developed a tool \cite{biker_tool} based on this approach. However, SO-based approaches \cite{rack, biker, biker_tool} only work for the APIs which are highly discussed in SO; SO may be slow to cover new APIs. Besides, there is a lack of security tool API related question answer in SO (discussed in Section 1). Due to these issues, we have not considered SO post for security tool API recommendation, rather focusing on API documentation.

Unlike the previous studies, Xie et al. \cite{verb_match} perform manual identification, categorization of functionality verb, and verb phrase patterns generation. Depending on the manual analysis authors use similarity measure to match functionality verb phrases in API description and query. This huge manual effort is both costly and time-consuming for diverse API documentation of different security tools. Unlike their approach, we perform data augmentation which enriches the corpus automatically with semantic variation of not only similar functionality verbs but also other Parts-Of-Speech (POS) tagged words as well. 

A set of studies \cite{mulapi, deepapi, portfolio, chan, swim} have used open source code repositories such as Github to recommend API usage and code snippet corresponding to a task. However, security tools used in a SOC include legacy, and proprietary tool and SOAR platforms are vendor and organizational application specific. The API usage in a SOAR platform is not freely accessible and open source. Unlike these works that rely on code repositories, our work relies on API documentation due to a lack of open source repositories of SOAR.

\vspace{-8pt}
\subsection{Advanced NLP and DL in API Recommendation}
Application of advanced NLP and deep learning approaches are observed in API recommendation \cite{slr, adaptive_code, deepapi} and artifact content analysis \cite{opinion_slr}. Researchers \cite{biker, api2vec} have adopted word embedding based (W2V-IDF) similarity score approach for API recommendation leveraging SO or GitHub. Gu et al. \cite{deepapi} use Recurrent Neural Network (RNN) to generate API call sequence for a natural language description from GitHub. Chen et al. \cite{chen2018neural} present a deep code search method called BVAE, which uses Variational AutoEncoders on code repositories. Ling et al. \cite{adaptive_code} propose AdaCS, which uses word embedding and RNN approach on code repositories for searching code. These approaches leverage external resources (e.g., SO, code repositories), which can only capture the most discussed APIs. Thus, for the large amount of less frequently used APIs, a user still have to consider documentation-based retrieval. Unlike these works, we do not rely on external resources. We use API documentation which covers all the APIs; that is why we assert that our approach does not suffer from inability of recommending less frequently used APIs. 

Growing interest is noticed in the adoption of FastText for different applications such as predicting issues type on GitHub and classifying software requirements \cite{fasttext_classify, fasttext_github}. FastText word embedding refers to the technique of learning distributed representation for words using sub-word information \cite{fastText_mikolov}. It uses skip gram to represent a word embedding vector as the sum of its constituent character n-gram vectors. In addition, several studies have shown the use FastText and it's capability for representing the Out-of-vocabulary (OOV) words \cite{concept_drift, polisis}. Thus, we use fastText to create the semantics-preserving word embedding (including OOV words), where similar word embeddings represent similar semantic words. 

The recent research trend is to provide various text augmentation approaches \cite{nlpaug, nlp_cloud, easyaug, contextual_data, wordnet_query}.
Text data augmentation refers to the method of synthesizing new data from the existing textual data which has gained popularity in NLP very recently from 2019. Various text augmentation approaches such as dictionary-based synonym or para-phrase replacement, context-dependent synonym incorporation, word occurrence statistics-based word insertion or replacement are available in the existing literature \cite{nlpaug, eda, niacin}. Nlpaug \cite{nlpaug}, niacin \cite{niacin}, and EDA \cite{eda} are popular libraries used for performing text augmentation.
Qiu et al. \cite{easyaug}, Kobayashi et al. \cite{contextual_data}, and Fadaee et al. \cite{machine_trans_aug} present the effectiveness of applying text augmentation in different classification models, NLP language model, and machine translation model, respectively. Lu et al. \cite{wordnet_query} have extended the query with synonyms from WordNet and used similarity measure to map query with source code identifiers of a code base. In contrast, we perform augmentation to mitigate the data availability constraint and to incorporate semantic variation in API descriptions reported in API documentation. We propose a learning-based framework to map query with augmented API descriptions, where augmentation is performed using WordNet synonyms and 35 other data augmentation techniques.

\begin{figure}[h]
  \centering
  \includegraphics[scale = 0.6]{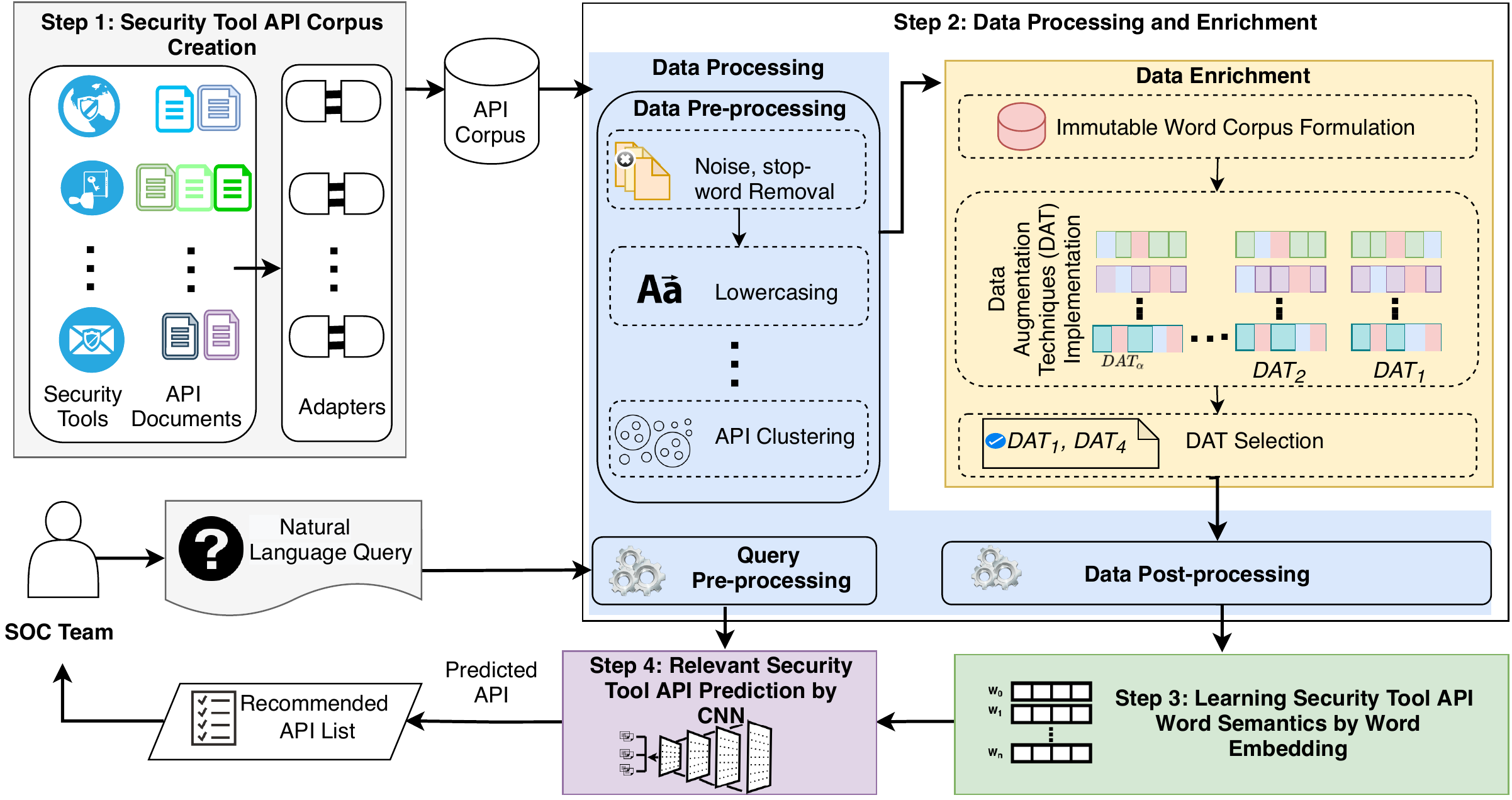}
  \caption{High-level overview of APIRO framework}
  \label{fig:framework}
  \vspace{-15pt}
\end{figure}

\section{APIRO Framework} \label{sec: framework}
We have proposed a framework, APIRO (\textbf{API} \textbf{R}ecommendation for security \textbf{O}rchestration, automation, and response) for automatically identifying and recommending suitable APIs of security tools. The proposed framework comprises of four steps as shown in Fig. \ref{fig:framework}: (i) security tool API corpus creation, (ii) data processing and enrichment, (iii) learning security tool API word semantics by word embedding, and (iv) relevant security tool API prediction by CNN. APIRO provides a ranked list of the relevant APIs with corresponding API information (Tool name, API, description, parameter, and return data) in response to queries regarding API. We describe the details of each phase of APIRO below.

\subsection{Security Tool API Corpus Creation}\label{subsec:api_data_collect}
\begin{figure}[ht]
  \centering
  \includegraphics[scale = 0.3]{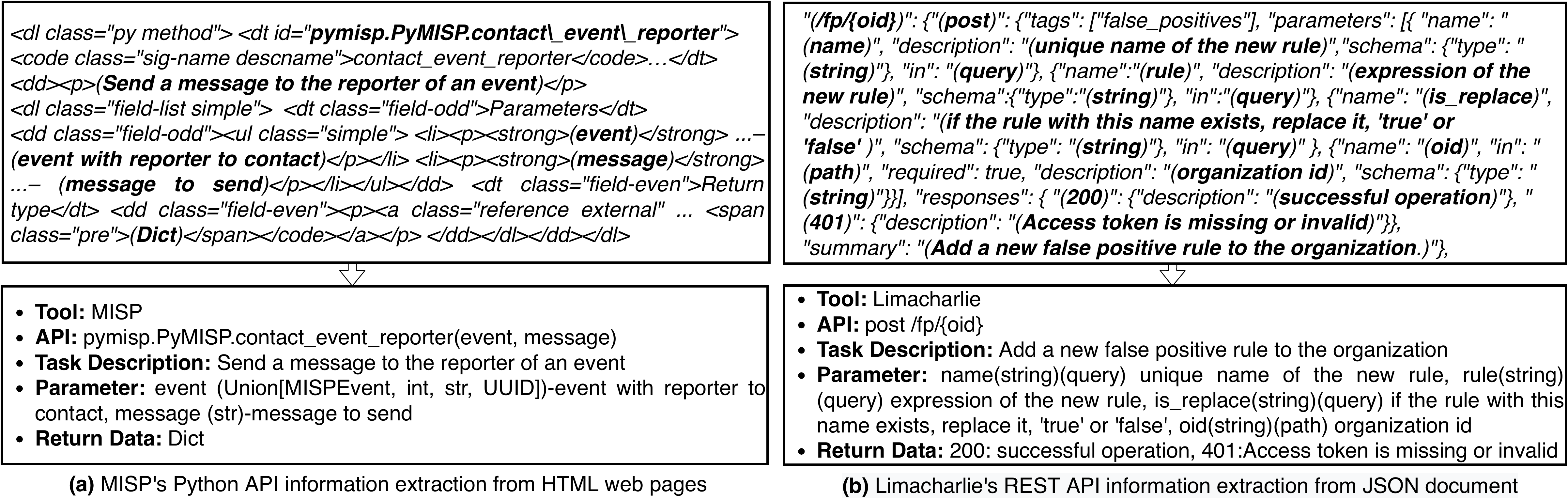}
  \caption{API information extraction example using different adapters}
  \label{fig:div_api_doc}
   \vspace{-10pt}
\end{figure}

APIRO has a security tool API corpus creation module (i.e., data collector) to gather API information of different security tools. We have analysed several security tools to show how these security tools may vary from each other to create security tool API corpus. Our analysis revealed five main attributes based on which these tools vary. These attributes are tool type, main functionalities, tool accessibility, API format, and API documentation source. Table \ref{tab:tool_att} shows the example values for these attributes. Diverse security tools can have diverse API documentation sources. For example, Fig. \ref{fig:div_api_doc} shows, while MISP’s Python API information is available in HTML web pages, Limacharlie REST API information is represented in JSON document. The task of collecting the security tool API information from diverse security tools, thus, require the creation of different adapters to scrape data for different security tools of interest. Hence, a list of adapters is exemplified in the security tool API corpus creation module in Fig. \ref{fig:framework}. New adapters are required to develop in `Security Tool API Corpus Creation’ module to include new data sources of different formats (e.g., YML and XML). The newly added adapters will incorporate the scripts of scraping data from different formats of document.

\begin{table}[ht]
\caption{Diversity of security tools based on different category attribute with example attribute values}
  \label{tab:tool_att}
\begin{tabular}{lp{11cm}}
\toprule
Tool category attribute & Example attribute values\\ 
\midrule
Tool type & EDR, IDS, Sand-boxing tool, Threat Intelligence Platform\\
Main functionalities & Real-time traffic analysis, Share IoC \& indicators database about malware, Packet logging, Analyze malicious files\\
Tool accessibility & Open-source, Commercial\\
Tool API & REST, Python, CLI Commands\\
API doc source & JSON Doc, HTML Doc \\
\bottomrule
\end{tabular}
 \vspace{-10pt}
\end{table}

Security tools' API documentations are considered as main data source of security tool API corpus creation module because they provide the details about APIs calls/commands, utilities, and tools \cite{api_doc_def}. A recent study by Michael et al. \cite{api_doc_survey} on 112 participants mentions API documentation as the first information source when developers need to solve problems. API information collected from API documentation include tool name, API, API description, parameters, and return data as shown in example API information extraction in Fig. \ref{fig:div_api_doc}. A security tool ($S^i$) can have multiple API documentation ($Doc^i_j$). For example,  Limacharlie has separate documentation for Python API  \cite{lima_python} and REST API \cite{lima_rest}. API information of a security tool, $S^i$ is collected from all API documents, $\cup^m_{j=1} Doc^i_j$, of a particular tool. Here, $j$=1, 2,..., $m$ and $m$ presents the number of available API documents. API information collected from $n$ number of security tools are further combined to create a unified API corpus, $C_{API}$ where

\vspace{-5pt}
\begin{equation}\label{eq:sec3}
    C_{API}=|\cup^m_{j=1} Doc^1_j| \cup |\cup^p_{j=1} Doc^2_j| \cup ... \cup |\cup^q_{j=1} Doc^n_j|
\end{equation}

\subsection{Data Processing and Enrichment}\label{subsec:processing-enrich}
APIRO's data processing and enrichment module has two components for processing and enriching data. Data processing component has three sub-components which are data pre-processing, data post-processing and query pre-processing. Processed and enriched data from this module will be used to create the word embedding model by the next module. Following we discuss each of these components:
\subsubsection{Data pre-processing}\label{subsec:pre-processing} 
\begin{figure}[h]
  \centering
  \includegraphics[scale=0.5]{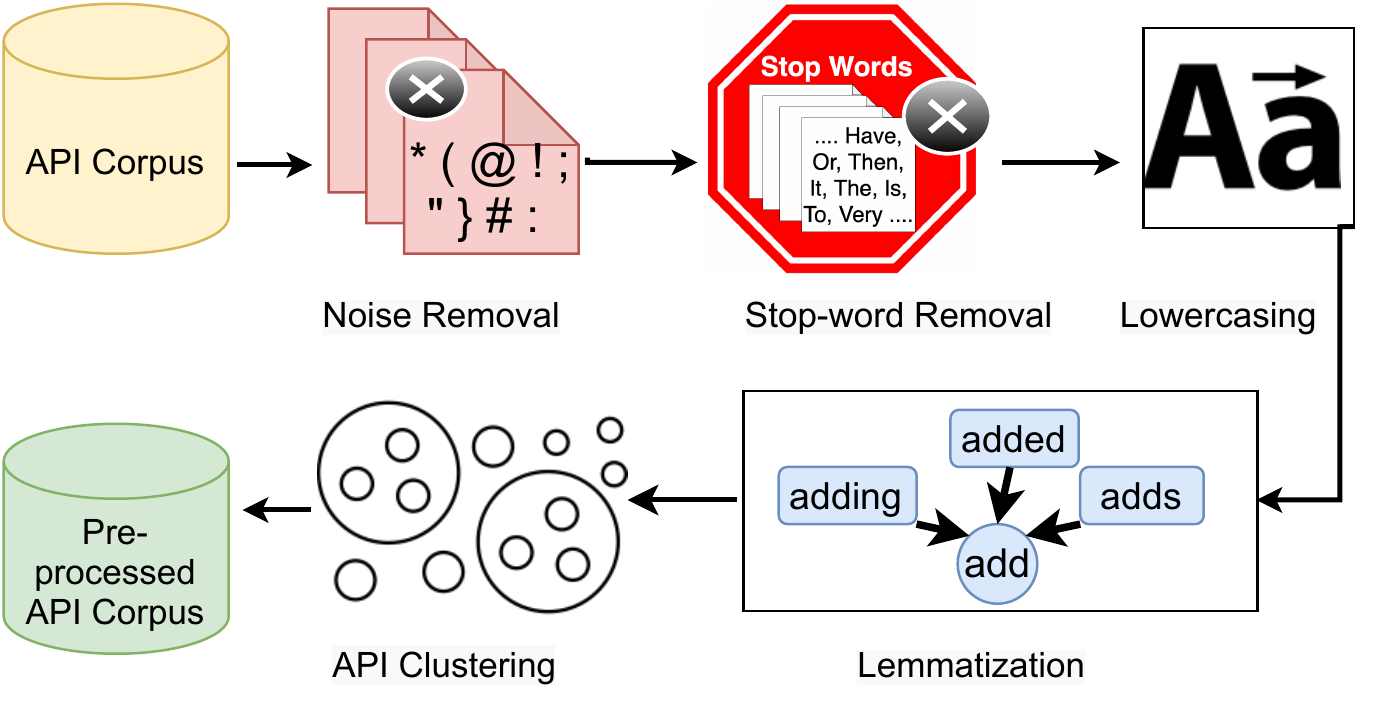}
  \caption{Data pre-processing steps in APIRO}
  \label{fig:pre-process}
   \vspace{-10pt}
\end{figure}

The data pre-processing sub-component automatically process the API descriptions of the unified API corpus ($C_{API}$) by following three steps, shown in Fig. \ref{fig:pre-process}, which are discussed below.

{\textbf{Noise removal}}\label{subsec: noise_remove}
purports to clean the API descriptions of $C_{API}$ corpus by removing all the noises (e.g., `.', `:') and keep only the alphanumerical values along with underscore and minus symbol. Underscore and minus are kept because removing underscore will create multiple sub-words from a single word (e.g., sharing\_group, misp\_entity) and minus is used to present arguments such as (e.g., -e, -t). 

{\textbf{Stop-word removal}}\label{subsec: stword_remove} is performed to remove the words  that appear frequently (e.g., a, is) but are not helpful to distinguish one API description from another. 

{\textbf{Lower-casing }}\label{subsec: lowercase}
maps words of distinct cases to the identical lower-case form. Training a word embedding model for diverse variation in input capitalization such as `MISP' vs. `misp' can give different vectors. A dataset with mixed-case occurrences (e.g., `MISP', `Misp', `misp') of a word can have inadequate support for neural networks to learn weights effectively in case of less common versions of a word.

{\textbf{Lemmatization}}\label{subsec: lemmatize}
is performed to eliminate different inflected forms of a word for mapping it to the root form (i.e., lemma). For example, ‘gets’ and 'getting' will be reduced to thier root form ‘get’. We use WordNet Lemmatizer that uses the WordNet \cite{wordnet} database to lookup lemmas of words. Wordnet is a large lexical database of English. Lemma is returned by vocabulary and morphological analysis of words in Wordnet to provide the base or dictionary form of a word.

{\textbf{API clustering}}\label{subsec: merge_api_method}
is proposed to combine multiple API in a specific security tool that has identical API description. By analysing security tool API information, we define four categories that are based on class, method, parameter, and representation. Some APIs with the same task descriptions differ from each other based on these four categories as shown in Table \ref{tab:duplicate_api}. Category 1 represents the APIs with the same method and representation. For example, Table \ref{tab:duplicate_api} shows two APIs of category 1 where \textit{pymisp.MISPObjectAttribute()} and \textit{pymisp.MISPAttribute()} classes have same method, \textit{hash\_values()}. Category 2 presents the APIs with the same class and representation. \textit{addUser()} and \textit{addUserPermission()} are under same class \textit{limacharlie.Manager.Manager()} with same task description, hence clustered under category 2. Category 3 presents the APIs with the same class, method, and representation. In this category, the main difference is the use of different parameters. Category 4 is for API with different representations and the same task description. For example, Limacharlie Python API \cite{lima_python} \textit{limacharlie.Sensor.Sensor.untag untag(tag)} and a Limacharlie REST API \cite{lima_rest} \textit{delete /{sid}/tags}' have identical task description `Remove a Tag from the Sensor.



\begin{table*}[thb]
\centering
\caption{Examples of API variation with identical description based on class, method, parameter, and representation}
\label{tab:duplicate_api}

\begin{threeparttable}
\resizebox{\textwidth}{!}{
\begin{tabular}{cccccl}
\toprule
Category & Class  & Method     & Parameter  & Representation & Example \\
\midrule
& & & & & API 1: pymisp.MISPObjectAttribute.hash\_values (algorithm=`sha512') \\
1 & $\times$ & $\checkmark$ & $\checkmark$ / $\times$ & $\checkmark$ & API 2: pymisp.MISPAttribute.hash\_values (algorithm=`sha512') \\
& & & & & Description: Compute the hash of every values for fast lookups                     \\
\hline
 & & & & &  API 1: limacharlie.Manager.Manager.addUser(email) \\
2 & $\checkmark$ & $\times$ & $\times$& $\checkmark$ &API 2: limacharlie.Manager.Manager.addUserPermission(email, permission) \\
& & & & & Description: Add a user to an organization.            \\
\hline
& & & &  & API 1: get  /tasks \\
3 & $\checkmark$ & $\checkmark$ &  $\times$ & $\checkmark$ & API 2: get   /tasks/{cmd} \\
& & & & & Description: Get usage information for commands that can be sent to sensors.    \\
\hline
 & & & & & API 1: limacharlie.Sensor.Sensor.untag 
untag(tag) \\
 4 & - & - & - & $\times$ & API 2: delete   /{sid}/tags \\
& & & & & Description: Remove a Tag from the Sensor.   \\ \hline
\end{tabular}}
\end{threeparttable}
\begin{tablenotes}
      \item\label{tnote:robots-r1} \footnotesize{$\checkmark$ represents same, $\times$  represents not same, -  represents not applicable}
    \end{tablenotes}
\end{table*}

The APIs with the same task description need to be merged in an API cluster considering the following two reasons:
\begin{itemize}
    \item If APIs with identical task description in a security tool are not clustered, a specific required API by a user may fall in lower rank in the recommendation list as there are several other APIs with the same task description.
    \item When a user queries about API for a specific task, recommending an API cluster will provide the user a comprehensive view of all the APIs which can provide the same intended task. 
\end{itemize}

An API cluster can be created automatically based on identical task description. An API cluster ($Cl_i$) is formed by incorporating all the APIs with identical description. Then, a cluster corpus ($C_{cl}$) is created incorporating all the API clusters ($Cl_i$). Finally, the API corpus ($C_{API}$) will be formed according to Equation \ref{eq:cls3}, where, $E_{cl}$ presents individual API entities of all the clusters.
\begin{equation}\label{eq:cls3}
    C_{API}=(C_{API} \cup C_{cl})-E_{cl}
\end{equation}


\subsubsection{\textbf{Data enrichment}}\label{subsubsec:data_aug}
\begin{figure}[ht]
  \centering
  \includegraphics[scale=.45]{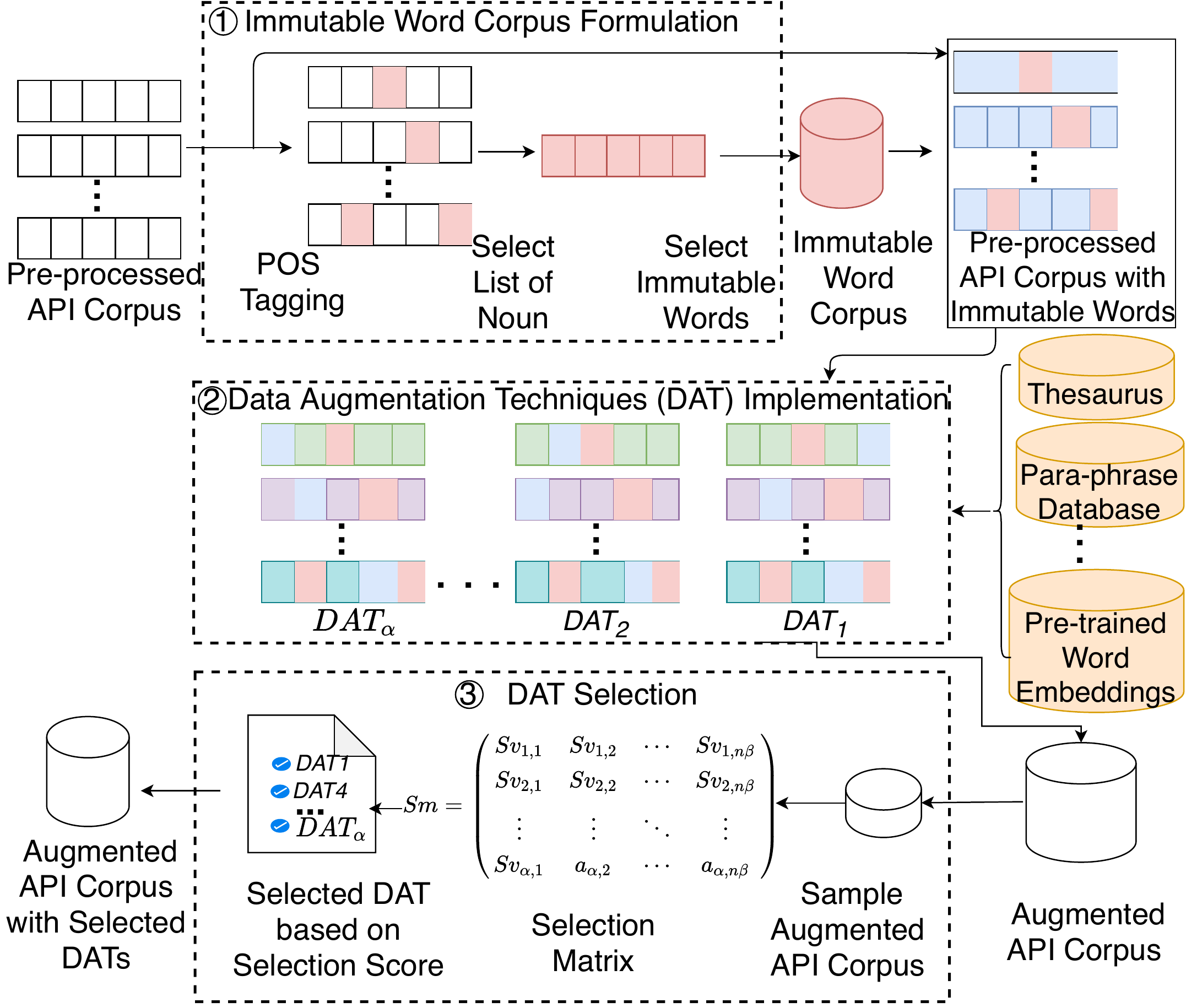}
  \caption{Data augmentation steps in APIRO}
  \label{fig:data_aug_flow}
   \vspace{-15pt}
\end{figure}

APIRO has a data enrichment (i.e., data augmenter) component that utilizes data augmentation techniques to synthesize new data using the security tools' API corpus gathered by the data collector module. We propose to apply augmentation techniques to mitigate the data availability constraint. Besides, $C_{API}$ corpus collected from the documentation does not reflect much semantic variation, i.e., an API with only one corresponding description that will not help to learn the possible semantic variations in query. The objective is to build enriched training data with semantic variations to improve the performance of the neural network model for recommending APIs to natural language queries. To enrich the security tool API corpus, as shown in Fig. \ref{fig:data_aug_flow}, the first step is to formulate an immutable word corpus for security tools, then implement different augmentation techniques for enriching the data set, and finally select the suitable augmentation techniques for security tool API corpus.

{\textbf{Immutable word corpus formulation:}}
Among various text augmentation techniques some augmentation techniques substitute words randomly with synonyms, paraphrases, semantic similar word from pre-trained word embeddings and contextual word embeddings. Applying text augmentation is challenging since meaning of words depends on the context and changing some words may change the context. Therefore, it is important to identify the immutable words from API descriptions that can not be augmented and need to keep unchanged. For example, API descriptions of $C_{API}$ corpus contain a set of words where substituting those words (i.e., an augmentation approach) may change the context or may provide semantically irrelevant words. Consider the API `\textit{./snort -no-interface-pidfile}’ that has a description \textit{`do not include the name of the interface in the PID file'}. Applying \textit{WordNet synonym substitution} replaces `PID' with `pelvic inflammatory disease'. However, in this context `PID' is referring to a file created in log directory after running Snort as a daemon \cite{snort_pid}. This substitution changes the context, thus not suitable for this particular API data. 

To mitigate this challenge, we propose to formulate an immutable-word corpus to be used in data augmentation. To formulate the immutable-word corpus, a Parts-Of-Speech (POS) tagged corpora is to be created from the pre-processed corpus. Our manual analysis of the POS tagged corpora (as detailed in Section \ref{subsec:methodology}) found that, words in the POS tags of the corpus such as ADP adposition (e.g., within, among), ADV adverb (e.g., immediately, alternatively), CONJ conjunction (e.g., either), DET determiner (e.g., every, another), NUM numeral (e.g., two, 2), PRT particle (e.g., back), PRON pronoun (e.g., us), VERB verb (send, remove) are usually changeable. However, various words under NOUN tag (e.g., syslog, webhook, xml) are immutable. Hence, the list of words under NOUN tag set needs to be inspected manually to select and create the immutable word corpus. 

\begin{figure*}[ht]
  \includegraphics[width=\textwidth, height=1.5in]{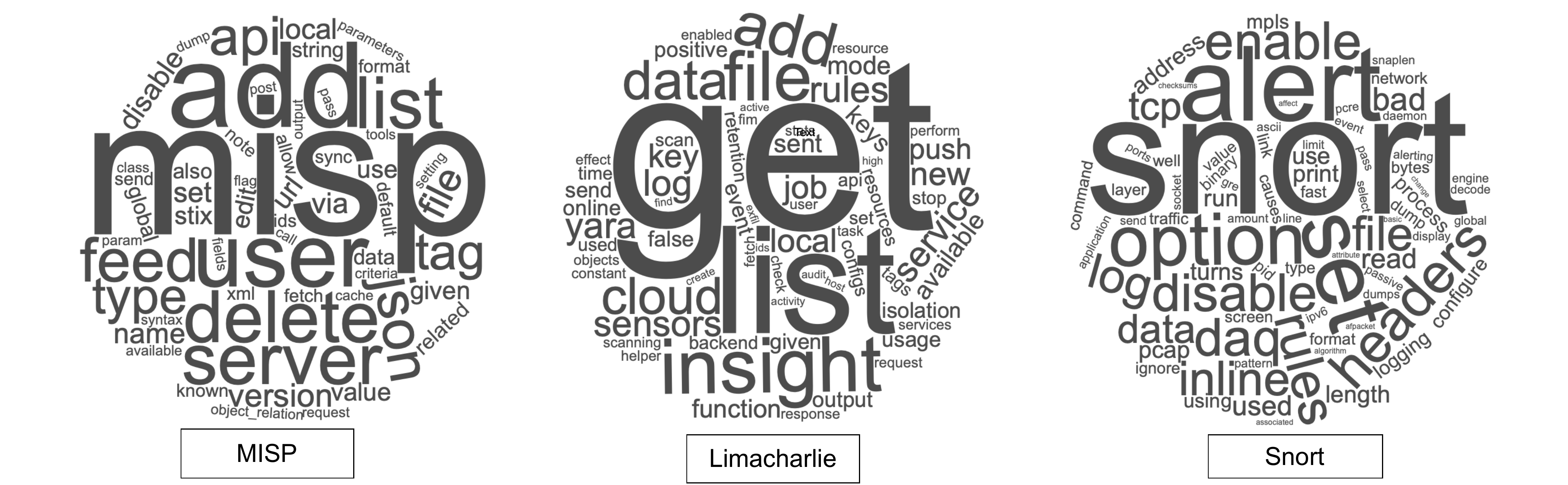}
  \caption{Word clouds of security tool API documentations}
  \label{fig:w_cloud}
\end{figure*}

Fig. \ref{fig:w_cloud} shows the word clouds of top-100 frequently-used words in API documentation of three different security tools. As shown in the word clouds, even though API documentation of each security tool uses their own tool-specific words (e.g., pcap for Snort, stix for MISP), at the same time they share some common words (e.g., api, json, cloud, python). This asserts while selecting immutable words for new security tools data, a portion of Nouns of the new tools tends to be common to the Noun list of existing security tools. Removing the common nouns reduces the number of Nouns to be analysed for manual inspection and selection of immutable words. Thus, each time a new tools data is added, the required immutable word selection efforts and time is reduced (detail analysis is given in Section \ref{subsec:methodology}).


\textbf{Data augmentation techniques implementation:} Various text augmentation approaches have been proposed in the existing research and new approaches are evolving \cite{nlpaug, nlp_cloud, easyaug}. We propose to apply a set of augmentation approaches (\textit{Aug}) as listed in Table \ref{tab:data_aug_rationale} to find the suitable ones for the gathered API corpus. Table \ref{tab:data_aug_rationale} shows the description and rationale for the usage of different augmentation approaches. To augment API descriptions using these augmentation approaches (\textit{Aug}), we aim to consider different actions (e.g., insert, substitute). Moreover, for augmenting data with the same \textit{Aug} and same action, we propose to use a wide variety of data sources, i.e., thesaurus and embedding (e.g., PPDB, GloVe Common Crawl). Hence, a set of $\alpha$ number of Data Augmentation Techniques $DAT$={$DAT_1$, $DAT_2$,..., $DAT_\alpha$} is created considering all combination of different \textit{Aug} approaches, each with diverse action and data sources. To implement any $DAT$, we need to incorporate the immutable word corpus, so that augmentation is performed without changing the immutable words. Implementation of $\alpha$ number of augmentation techniques creates the augmented API corpus ($C^{aug}_{API}$), which also includes the pre-processed original API data from documentation. We provide the details of the data augmentation techniques ($DAT$) with corresponding $Aug$ approaches, actions, data source with details, and the data source availability link in Appendix (A.2). 

\begin{table*}[ht]
\centering
\caption{Data augmentation approaches with corresponding rationale for usage}
\label{tab:data_aug_rationale}
\footnotesize
\begin{tabularx}{\textwidth}{p{3cm}p{3.9cm}p{7.5cm}}
\toprule
{ \textbf{Aug Approaches}} & {\textbf{ Description}} & {\textbf{ Rationale for usage}} \\
\midrule
{ \textit{$Aug_1$.}RandomWordAug} & { Swap/delete words randomly} & { Incorporate short and unordered API description variation} \\
{\textit{$Aug_2$.}SpellingAug} & { Use words with spelling mistake} & { Incorporate spelling mistakes in API description} \\
{\textit{$Aug_3$.}SplitAug} & { Split words randomly} & { Include typing errors (i.e., extra space splitting an word) in API description} \\
{\textit{$Aug_4$.}SynonymAug} & { Use synonym or paraphrases} & { Incorporate synonym or paraphrases in API description} \\
{\textit{$Aug_5$.}TfIdfAug} & { Use words based on TF-IDF statistics} & { Use word frequency statistics of API documentation to augment description} \\
{\textit{$Aug_6$.}WordEmbsAug} & { Leverage pre-trained word embeddings} & { Incorporate semantically similar words in API description } \\
{\textit{$Aug_7$.}ContextualWordEmbsAug} & { Leverage contextual word embeddings} & { Include contextually similar words in API description}\\
\hline
\end{tabularx}
\vspace{-10pt}
\end{table*}

{\textbf{Data augmentation techniques selection:}}\label{subsubsec:data_aug_select} 
All data augmentation techniques do not preserve the semantics and context of the text. We need to identify and select the suitable augmentation techniques for security tools' API data augmentation. We propose to build a sample corpus from the augmented API corpus for selecting suitable data augmentation techniques for the enrichment of security tool API data. The sample corpus is built by randomly sampling API data from each security tool for all the applied augmentation techniques. The size of the sample corpus should be statistically significant (commonly used sample size with 95\% confidence level and 5\% error \cite{sample_tech}) to the augmented API corpus. Then, the sample corpus needs to be inspected to label the augmented data with selection value $S_v$ as semantically relevant or irrelevant. $S_v = 1$ should be given if the augmented data is semantically similar or semantically related to the original data otherwise $S_v = 0$.

Equation \ref{eq:select} is designed  to calculate the selection score for each of the augmentation techniques, considering we have $\alpha$ number of data augmentation techniques. In Equation \ref{eq:select}, the sum of the selection value of all data samples for a specific augmentation technique ($DAT_\alpha$) is divided by the total number of API samples. Here, $Sv_{\alpha i}$ refers to the selection value of semantic relevance of an augmented API for augmentation technique $DAT_\alpha$,  $\beta$ refers to the number of APIs selected per tool and $n$ refers to the number of tools. 

\begin{equation} \label{eq:select}
            S_{score}{(DAT_{\alpha})}= \dfrac{\sum_{i=1}^{n\times \beta} Sv_{\alpha i}}{n\times \beta}\times100
\end{equation}



After calculating selection scores of each $DAT$, we have to calculate the mean selection score ($M_{score}$) by calculating mean of selection scores of each $DAT$. Then, we remove augmented data from $C^{aug}_{API}$ corpus, which are generated from augmentation techniques with selection score less or equal to the $M_{score}$. Hence, after this step, the augmented corpus $C^{aug}_{API}$ includes the original data and data generated from the selected augmentation techniques with corresponding API, parameter, and return value.

\subsubsection{\textbf{Data post-processing}}\label{subsec:postprocess}
This sub-component performs post-processing to clean the API description of augmented corpus $C^{aug}_{API}$ that may include stop-words, noise, or upper-case words. Post-processing module performs four data processing techniques that are noise removal, stop-word removal, lower-casing, and lemmatization as described in Section \ref{subsec:pre-processing}. 
 
\subsubsection{\textbf{Query pre-processing}}\label{subsec:queryprocess}
Queries about APIs are processed in this sub-component. Pre-processing steps noise removal, stop-word removal, lower-casing, and lemmatization as described in Section \ref{subsec:pre-processing} are applied to process a query. For example, consider a query `How do I add False Positive rule?', after pre-processing the query will be `add false positive rule'.

\subsection{Learning Security Tool API Word Semantics by Word Embedding}\label{subsec:word-embded}

Answering to free-form queries require to learn the variation of semantically similar words. We learn the variation of semantically similar words using fastText \cite{fastText_mikolov} word embedding on the API descriptions of augmented corpus $C^{aug}_{API}$. These semantic preserving word embedding from this module will be used in the next module while training the CNN model for API recommendation. The aim of APIRO is to answer free form query, which can contain variations of words. For example, a query may have synonyms or paraphrases. Consider a query `how to insert new false positive rule?' for the API description `add new false positive rule organization'. Replacing the word `add' with `insert', such variation which is unseen to the trained CNN model results in OOV problem. The occurrence of Out-Of-Vocabulary (OOV) word is reported as a common phenomenon in the existing literature in query-based API recommendation \cite{biker, rack} and privacy policy recommendation \cite{polisis}. Since APIRO aims to support free-form natural language queries, we cannot enforce users’ word usage and OOV words can be part of any query. Besides, spelling mistakes are common in queries, which can lead to OOV words. Kaibo et al.  \cite{query_reform_so} showed that 19.1\% of their analyzed SO queries had spelling and syntax issues. A CNN model can be confined to the training data without knowing the semantic variation of similar words. Distributed word representations are able to hold word semantics as similar word embeddings represent similar semantic words \cite{word2vec_mikolov, fastText_mikolov}. Thus, we propose to learn distributed word representations using word embedding models (e.g., fastText) from the descriptions of $C^{aug}_{API}$.

Moreover, free form query can contain variations of words, which are not available in the vocabulary such as: 
\begin{itemize} \itemsep -0.5 em
    \item User query often have out-of-vocabulary (OOV) words \cite{rack}, that are not present in the training corpus. For example, consider words such as `misp' and `attribute' are present in the vocabulary but a query contains a compound word `misp\_attribute', which is absent in the vocabulary. 
    \item Query can have word that share same radical (base form) with words in the API corpus such as (`\textbf{IDS}', `N\textbf{IDS}', `H\textbf{IDS}') and (`\textbf{log}', `\textbf{log}file').
    \item Query often has spelling or typing error \cite{polisis} such as with added space in the word `taxonomy' can be mistakenly typed as `ta xonomy' in a query.
\end{itemize}
APRIO needs to create embedding for the above variations in a query that have not been seen in the training corpus. However, word2vec \cite{word2vec_mikolov} embedding model can not represent the OOV words, as word2vec only considers context words for creating embedding. Thus, we adopt fastText \cite{fastText_mikolov} to learn the embedding matrix $E$, which can represent the OOV words. FastText allows training vectors for subwords (i.e., character n-grams) in addition to words to represent the OOV words.

\begin{figure}[hbt]
  \centering
  \includegraphics[scale=.6]{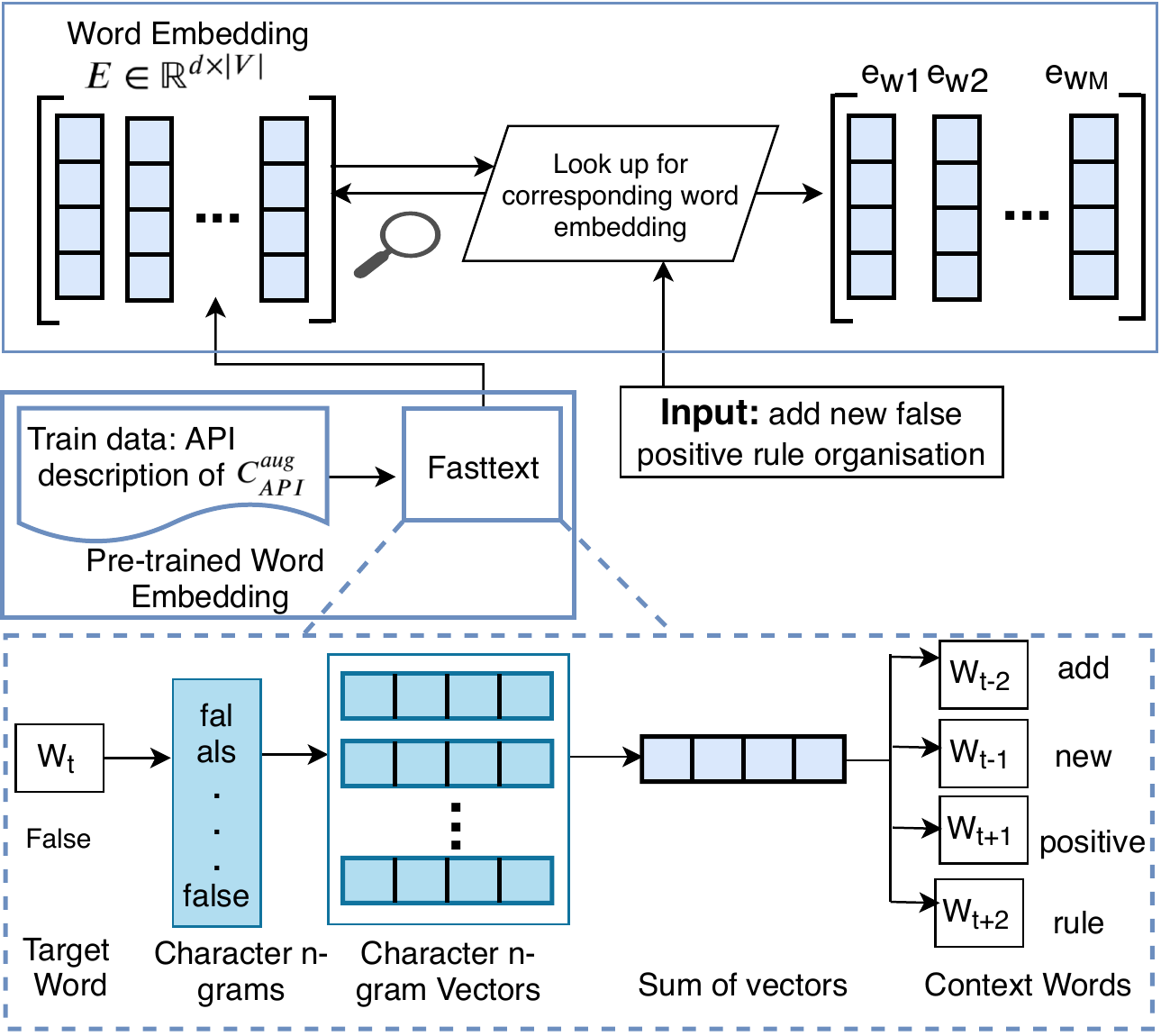}
  \caption{Word embedding in APIRO}
  \label{fig:wemb}
   \vspace{-10pt}
\end{figure}
As shown in Fig. \ref{fig:wemb}, we encode word embedding matrix, $E$ by corresponding column vectors. The embedding matrix $E$ is to be learnt from the API descriptions of $C^{aug}_{API}$ using fastText. Consider, $V$ is the vocabulary of words for API descriptions of $C^{aug}_{API}$. $E\in$ $\mathbb{R}^{d \times|V|}$, where $d$ and $|V|$ denote word embeddings dimension and vocabulary size, respectively. Fig. \ref{fig:wemb} shows fastText embedding approach to build $E$ with an example. Consider `false' is a target word for the pre-processed API description `add new false positive rule organization'. Firstly, embedding for a target word ($w_t$=`false') is to be calculated by the summation of vectors for the character n-grams (e.g., `fal', `als') and the whole word (`false') itself. Next, the embedding vectors are updated to bring actual context words ($w_{t-2}$=`add', $w_{t-1}$=`new', $w_{t+1}$=`positive', $w_{t+2}$=`rule') closer to the target word. The fastText word embedding for the $t$-th word from the vocabulary $V$ corresponds to column $E_t \in$ $\mathbb{R}^d$. To get the word embedding $e_w$ of any Input word $w$, we look up $E$ with the word $w$ such as $e_w=Ev^w$. Here, $v^w$ denotes one-hot vector with size $|V|$ having 1 at only index $w$ and 0 in all other positions. Hence, for recommending to free form natural language query, we can create the embedding of the OOV words, words with spelling mistakes, or words sharing same radical in query, by combining the vectors of their constituent subwords using fastText model.


\subsection{Relevant Security Tool API Prediction by CNN}\label{subsec:cnn}

Answering to free-form queries require a recommendation model with better generalization capabilities, which we achieve by using CNN with integrated pre-trained word embedding on security tool API data. APIRO is proposed to be a unified framework which includes data from diverse security tools API documentation. Even though API documentation of each security tool uses their own tool-specific words (e.g., pcap for Snort, stix for MISP), at the same time they share some common words (e.g., get, organization), as shown in the word clouds of Fig. \ref{fig:w_cloud}. Since APIRO incorporates data from various security tools, APIRO needs to learn the different contexts for the same common word for different security tools. For example, APIRO needs to learn different contexts for the common word `get’ for different security tools such as `get pcap’ for Snort and `get stix’ for MISP. Hence, we need to learn the syntactic and semantic features from API descriptions for API recommendation, which we can learn automatically using the convolutional filters of the CNN model \cite{symantic_cnn}. Word embedding creates similar word vectors for mapping similar words from different security tools. Word embedding-based CNN helps to learn these different contexts of a word by using convolution filters of different sizes to create corresponding features in APIRO.

Other models such as RNN are also popular to learn the textual features. However, CNN is more suitable for APIRO due to the short description of APIs (e.g., the average length of API description of Limacharlie, MISP, and Snort are 11.15). RNN based models are more suitable to learn the sequence of data to understand the next sequence. For this, RNN-based models encode the input tokens sequentially and operations in RNN based network structures can not be parallelized, which results in low efficiency. A set of existing studies have supported CNN over RNN for the classification of textual data \cite{rnn_over_cnn1, rnn_over_cnn2}. An existing study also suggests that a combination of convolutional filters in a single convolutional layer achieves comparable good performance \cite{cnn_sentence}. Moreover, CNN does not require knowledge of the linguistic structure of the corpus \cite{character}. It explores the richness of the pre-trained word embedding models.

\begin{figure}
  \centering
  \includegraphics[scale=.5]{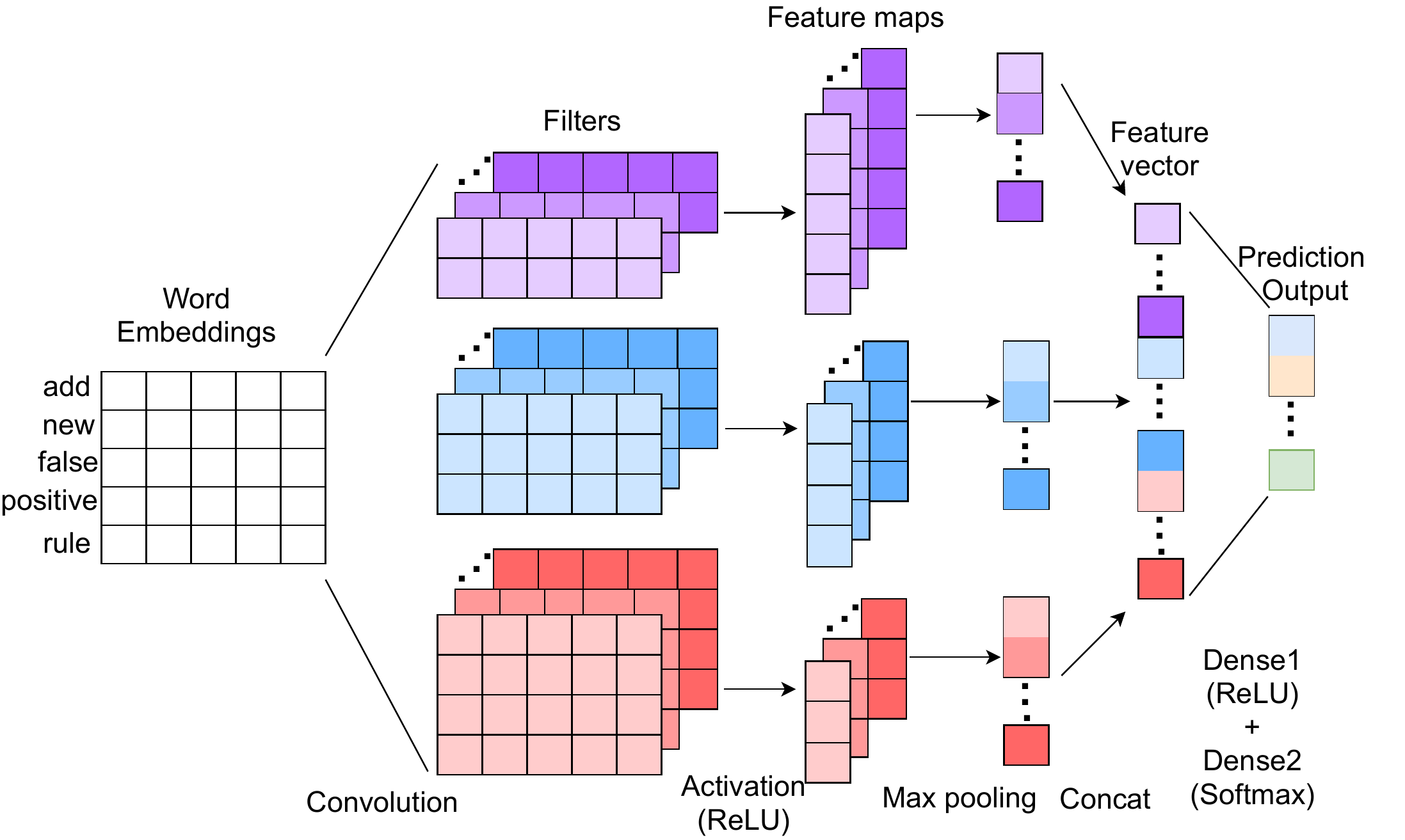}
  \caption{An illustrative example of convolutional neural network in APIRO}
  \label{fig:cnn}
   \vspace{-10pt}
\end{figure}

Fig. \ref{fig:cnn} represents the architecture of our word-level CNN. Considering an input text with $M$ words ({$w_1$ , $w_2$ , ..., $w_M$}), we look up the word embedding matrix $E$ (Fig. \ref{fig:wemb}) to obtain the sequence of word embeddings ({$e_{w_1}$ , $e_{w_2}$ , ..., $e_{w_M}$}). As illustrated in Fig. \ref{fig:wemb}, each $d$-dimensional column vector of word embedding matrix $E$ corresponds to a specific word. As shown in Fig. \ref{fig:cnn}, input of CNN model is the sequence of word embeddings, which passes through a Convolutional layer. We select the width of the convolution filters equal to the dimensionality of the word embeddings (i.e., $d$) in our CNN. We can consider varied height ($h$) of the convolution filter. Here $h$ (i.e., height or window size) represents the number of words considered in a convolution operation. 


A convolution operation involves applying a non-linear activation function which is Rectified Linear Unit (ReLU) \cite{cnn_sentence} over filter matrix $F \in$ $\mathbb{R}^{h\times d}$ with $h\times d$ dimensions. For input sequence {$w_1$, $w_2$, ..., $w_M$}, the convolution operation is repeatedly performed to each word window of different size ($h$). Feature map is produced by the convolution operations for a specific filter. We apply $N$ number of filters to the same word window of size $h$ for the extraction of different feature maps. Here value of $N$ is selected through hyper-parameter tuning. Then, we apply max-pooling operation to the feature maps for extracting the maximum feature map value. By concatenating the resultant feature vectors from max-pooling layer a single feature vector is created. This feature vector then passes through the first dense layer that is a fully-connected layer including ReLU activation function. Finally, through the second dense layer using softmax operation, the prediction probabilities are achieved. We use Sparse\_categorical\_crossentropy loss \cite{sparse_cat_loss} and adam optimizer \cite{sentiment_cnn} as the objective function of our model and training optimization algorithm, respectively. Hence, given a query $Q$, APIRO recommends the top-k relevant API using the top $k$ prediction probabilities provided by the CNN model.

\section{Experimental Setting and Implementation} \label{sec: evaluation}
This section presents our experiment design and implementation to evaluate APIRO. First, we propose our research questions. Then, we describe the dataset, implementation, baseline, and evaluation metrics. 

\subsection{Research Questions}
The experimental evaluation of APIRO was motivated by the following three research questions:

\textbf{RQ1:} Are data augmentation techniques suitable for API recommendation for natural language query?

\textbf{RQ2:} How does APIRO perform compared to the baseline method for security tool API recommendation?

\textbf{RQ3:} How effective is APIRO for answering free-form natural language query?

To mitigate data availability constraint, we aimed to evaluate the applicability of the data augmentation techniques using RQ1. Besides, RQ2 addressed the data heterogeneity challenge by evaluating the performance of proposed unified framework APIRO and comparing it to the state-of-the-art W2V-IDF approach. Moreover, to address the semantic variation challenge, we wanted to evaluate the effectiveness of APIRO in answering different free-form natural language query categories in RQ3.



\subsection{Dataset}\label{subsec:data_collection}

For our experiment, we created security tool API corpus using MISP, Limacharlie, and Snort, which are the three widely used security tools in the existing literature \cite{sec_orch_caise, islam2019ontology}. We selected these tools as according to Orange Cyberdefence \cite{orange_cyber} and the existing literature \cite{sec_orch_caise, islam2019ontology}, a common use case in SOAR platform of organizations is the use of Threat Intelligence Platform (TIP) (e.g., MISP) for the search of Indicators of Compromise (IoC). Then, to identify whether an indicator has been observed on endpoint or on network, the organizations use EDR tool (e.g., Limacharlie) and NIPS tool (e.g., Snort). As reported in Table \ref{tab:tool_comp}, these security tools support very distinct functionalities for ensuring security, and are also diverse in other category attributes such as tool API (e.g., REST, Python), API data source (e.g., JSON doc, HTML doc), and tool accessibility. The effectiveness of APIRO framework for API recommendation can be evaluated using API data of these tools in varied data settings. We collected available API information (i.e., API, description, parameter, and return data) from corresponding API documentation of these security tools to build dataset for our experiment.


\begin{table*}[ht]
\centering
\caption{Variety of security tools based on different category attributes (e.g., tool type, functionalities)}
\label{tab:tool_comp}

\begin{tabularx}{\textwidth}{lXXX}
\toprule
Tool name & Limacharlie & MISP & Snort \\ 
\midrule
Tool type & EDR & TIP & NIPS \\\hline
{\specialcell{Main\\ functionalities}} & User can write Detection \& Response Rules (DRR) & IoC database about malware, incidents, attackers \& intelligence & Real-time traffic analysis \& packet logging \\
 & Monitor any file type by DRR. & Data sharing functionality & Content searching \& matching \\
 & Any executable can be sent \& run in endpoint & Auto correlation of attributes \& indicators from malware & Detect a variety of attacks \& probes, analyze protocol  \\\hline
Tool accessibility & Commercial & Open source & Open source \\\hline
Tool API & REST, Python, CLI Commands & REST, Python & CLI Commands\\\hline
\specialcell{API doc source} & JSON Doc, HTML Doc & HTML Doc & HTML Doc (Semi-structured)\\

\bottomrule

\end{tabularx}
 \vspace{-10pt}
\end{table*}

\textbf{Limacharlie} is an EDR tool that performs real-time security attack detection on endpoints. We used Beautiful soup\footnote{https://www.crummy.com/software/BeautifulSoup/bs4/doc/} Python library to parse the HTML pages from Python API \cite{lima_python} and Sensor Commands \cite{lima_sensor_com} documentations of Limacharlie. For scraping the JSON data from Limacharlie REST API document \cite{lima_rest}, Python built-in JSON package was used. 

\textbf{MISP} is a widely adopted tool that supports collaborative-knowledge-sharing about malware and threats that help in attack detection. We followed a similar method to scrape  Limacharlie Python API for MISP. Data was scraped from PyMISP-Python Library \cite{PyMISP_apidoc}, PyMISP Python API \cite{PyMISP_lib}, and MISP Automation API \cite{misp_automation_api} HTML documents using Beautiful soup.
Examples of security tool API data extraction from HTML and JSON document are already shown in Fig. \ref{fig:div_api_doc}.



\textbf{Snort} is a widely used NIPS, which performs traffic analysis and logs packet in real-time. We collected data from Snort documentation \cite{snort_usermanual}, which is a semi-structured HTML document that hinders the automated scraping of API data. In the documentation \cite{snort_usermanual}, API syntax is text under \textit{<pre>} tag and description is under (\textit{<p>}) tag. Unfortunately, same \textit{<p>} tag can contain description of multiple APIs/commands. For example, in the following description, the later sentence describes another command in the same \textit{<p>} tag. 
\newline
``\textit{\textbf{<pre>}./snort -vd \textbf{</pre>}} \textbf{<p>} \textit{This instructs Snort to display the packet data as well as the headers. If you want an even more descriptive display, showing the data link layer headers, do this}: \textbf{</p>}.''
\newline 
Even in many cases the section or subsection header title (<h3>) can be an API description such as `\textit{\textbf{<h3>} Read a single pcap \textbf{</h3><pre>}\$snort -r foo.pcap \$snort -- pcap- single=foo.pcap \textbf{</pre>}}'. Hence, instead of automated tag-based API data extraction, it demands manual inspection for collecting API and corresponding data. Since manual data collection is both time and effort-intensive, a sample subset (first 41 pages out of 269 pages) was included in our corpus, which included the basic modes (Sniffer, Packet Logger, and NIDS) and basic functionalities (e.g., packet acquisition, reading pcap files) of Snort. 

\begin{table}[h]
\centering
\caption{Number of APIs, total word count, and mean word count in API description of collected security tool API data}
\label{tab:api_statistics}
\scalebox{0.9}{
\begin{tabular}{|l|l|c|c|c|}
\hline
\specialcell{\textbf{Security tool}}               & \textbf{Data source }            & \multicolumn{1}{l|}{\specialcell{\textbf{Num of APIs}}} & \multicolumn{1}{l|}{\specialcell{\textbf{Num of words in descriptions}}} & \multicolumn{1}{l|}{\specialcell{\textbf{Mean word count}}} \\ \hline
{\specialcell{Limacharlie}} & Python API              & 146                                & {2395}                             & {8.81}                                   \\ \cline{2-3}
                             & REST API (Swagger API)  & 84                                 &                                                   &                                                         \\ \cline{2-3}
                             & Sensor Commands         & 42                                 &                                                   &                                                         \\ \hline
{MISP}        & Automation and MISP API & 66                                 & {4811}                             & {13.75}                                  \\ \cline{2-3}
                             & PyMISP Python Library   & 20                                 &                                                   &                                                         \\ \cline{2-3}
                             & PyMISP Python API       & 312                                &                                                   &                                                         \\ \hline
Snort                        & Snort (Upto 2.2)        & 145                                & 1577                                              & 10.88                                                   \\ \hline
Total                        &                         & 815                                & 8783                                              & 11.15                                                   \\ \hline
\end{tabular}
}
\vspace{-10pt}
\end{table}


We built a unified security tool API corpus ($C_{API}$) according to Equation \ref{eq:cls3} by combining data collected from Limacharlie, MISP, and Snort. The dataset (i.e., API corpus) contains 815 APIs as shown in Table \ref{tab:api_statistics}. For API descriptions, the total word count is 8783 and the mean word count in an API description is 11. Since APIRO considers the official API documentation of different security tools as input, the API descriptions in the official documentation are presented in well-structured natural language format.

\subsection{Implementation}\label{subsec:methodology}
The experiment environment we used was the Phoenix\footnote{Phoenix HPC service is super-computing resources provided by the University of Adelaide}$^,$\footnote{https://www.adelaide.edu.au/technology/research/high-performance-computing/phoenix-hpc\#technical-details} High-Performance Computer (HPC) Service. Phoenix supercomputer hardware is a Lenovo NeXtScale system (Intel X86-64) consisting of 260 nodes, where each node has multiple CPU cores. We used one node with 10 CPU cores and 10 GB dedicated RAM for our experimental environment.

We used  NLTK's \cite{nltk} English stop-word list, lower-casing, and WordNet-based lemmatization to remove noise, stop-words, and lemmatize API description of dataset. We implemented a Python script to create clusters of APIs which differed based on our defined categories (described in Section \ref{subsec:pre-processing}) but had the same API description. API clustering created 44 API clusters by merging 199 APIs as shown in Table \ref{tab:duplicate_api_number}. After pre-processing, the security tool API corpus includes 660 unique API descriptions of APIs and API clusters.

\begin{table}[ht]
\centering
\caption{Number of API and API clusters in the corpus based on defined categories (showed in Table \ref{tab:duplicate_api})}
\label{tab:duplicate_api_number}
\begin{threeparttable}
\scalebox{0.9}{
\begin{tabular}{ccc}
\toprule
Category  & Num of API & Num of API clusters\\
\midrule
1     &    167 & 29          \\ 
2     &   10 & 5          \\ 
3     &     7 & 3          \\
4     & 15 & 7          \\ \bottomrule
\end{tabular}%
}
 \end{threeparttable}
\vspace{-10pt}
\end{table}

To build the immutable word corpus, we used Natural Language Toolkit (NLTK) POS tagger and Universal POS Tagset \cite{postag} that returned a POS tagged corpora of API descriptions of our dataset. The  NLTK POS-tagger is a robust open-source POS tagging library for English, which is implemented in Python. The accuracy of the NLTK POS-tagger across different corpus had been widely studied and showed comparative good performance in the existing literature \cite{nltk_pos_comp1, nltk_pos_comp2}. 

Table \ref{tab:imm_seletion} shows the statistics related to the selection of immutable words, which shows the number of Nouns in Limacharlie, MISP, and Snort are 270, 287, and 279, respectively. We created a ‘Noun List', ($N_{agg}$) where we aggregated the already analyzed unique Nouns. While building the immutable corpus for any new security tools, we only considered the ones that are not in $N_{agg}$. For instance, if we first considered Limacharlie, after labelling it's 270 Nouns we included it in $N_{agg}$. For the next tool, MISP, instead of labelling all it's 287 Nouns, we removed the repeated Nouns that are already in $N_{agg}$, which gave us 206 Nouns to label. If we next considered Snort, for Snort the number of Nouns to label is reduced by 37.6\% (as shown in Table \ref{tab:imm_seletion}). It shows that with the increase of security tools, the number of words to label as immutable decreases; showing the scalability of our approach in building immutable word corpus.

The first two authors independently analyzed and labelled the nouns of each security tool as 1 if it is immutable and 0 if it is not immutable. The Cohen’s Kappa value is 0.75, which indicates moderate agreement \cite{cohen_kappa} between the annotators. Besides, the agreement rate is 89.2\%. The disagreements were resolved through discussion to reach on agreement for mitigating the error possibility. 
Table \ref{tab:imm_seletion} further shows that on average 11 minutes were required by the two annotators for the manual labelling of each security tool's immutable words. The immutable corpus is provided in Appendix (A.1).

\begin{table}[]
\centering
\footnotesize
\caption{Immutable word selection statistics from different security tool API documentation}
\label{tab:imm_seletion}
\begin{tabularx}{\textwidth}{XXXXXXXXXX}\hline
Tool & Total Noun count & Common Noun count with $N_{agg}$ & Noun analysed manually & Reduction rate of Noun for analysis & Immutable word count & Immutable word to Noun ratio &  Common Immutable word count & Annotator1 (time in minutes) & Annotator2 (time in minutes) \\ \hline
1.Limacharlie & 270  & { -} & { 270} & { -} & 44 & { 16.3\%} &  { -} & { 14} & { 15} \\
{ 2. MISP} & 287  & { 81} & { 206} & { 28.2\%} &62 & { 21.6\%}  & { 13} & { 10} & { 8} \\
{ 3. Snort} & 279 & { 105} & { 174} & { 37.6\%} & 81 & { 29.0\%} &  { 10} & { 9} & { 10}\\\hline
\end{tabularx}
\vspace{-13pt}
\end{table}

We applied 36 data augmentation techniques. The variation of these techniques are discussed in experimental results (Section \ref{subsec: rq1}) and listed in Fig. \ref{fig:dataaug_select}. Detailed description of these augmentation techniques are given in Appendix (A.2). Since word replacement by antonym is not a semantically invariant transformation \cite{nlp_cloud}, we skipped this technique. As we lower-cased API description in the pre-processing step, so cased data sources were not considered for data augmentation (e.g., Common Crawl (2.2M vocabulary, cased), bert-base-cased). We used Nlpaug library \cite{nlpaug} to perform data augmentation. We loaded the stop-word list with our `immutable-word corpus' and kept the default values for the other parameters. After augmenting 660 distinct API descriptions of the security tool API corpus, the augmented dataset had 24420 different API descriptions.




Then, we created a sample augmented API corpus of 555 API data (with 95\% confidence level and 4.11\% error), which has less error than the state-of-the-art sample size with 95\% confidence and 5\% error (i.e., requires 379 API data) \cite{sample_tech}. We created the sample augmented API corpus for labelling by randomly selecting 5 augmented APIs data from each of the three tools with all the corresponding augmented data for those 15 APIs. Considering Equation \ref{eq:select}, for our sampled augmented API $\beta=5$ and $n=3$, thus $n\times \beta$ gives us 15 APIs. Thus for each of the 36 augmented techniques, we have 15 augmented API descriptions. We  labelled them as 1 if augmented description is semantically similar or semantically related to the original description, otherwise we labelled as 0 (semantically irrelevant). For example, the API with the original description ‘Add a new false positive rule to the organization’, we labelled all the 36 augmented description for this API description as 1 or 0. The first two authors of the paper labeled the sample augmented API corpus independently. Any disagreements were resolved through discussion to reach on agreement for mitigating
the error possibility. We used Cohen’s Kappa \cite{cohen_kappa} for calculating the agreement between the annotators. The Cohen’s Kappa value is 0.66, which indicates moderate agreement \cite{cohen_kappa} between the annotators. Besides, the agreement rate is 83.3\%. After the manual annotation disagreements were resolved through discussion to reach on agreement for mitigating the error possibility.

We calculated selection score of each $DAT$ using Equation \ref{eq:select}. Then, we calculated the mean value of these selection scores which is 53.5. Out of 36 augmentation techniques 19 were selected, which had the selection score greater than the mean selection score 53.5. Section \ref{subsec: rq1} provides further in-depth analysis of why a certain augmentation technique was selected or not selected. We built an augmented API corpus with the selected augmentation techniques. Noise removal, stop-word removal, lower-casing, and lemmatization were performed on the augmented data using NLTK.

We implemented fastText model using Gensim \cite{gensim} toolkit to learn the semantic meaning of different words that return security tool API-specific embedding model. Gensim \cite{gensim} is an open-source Python-based robust toolkit for vector space modeling and topic modeling. We used Skip-gram model to learn the word vectors in fastText approach, that is useful to predict the neighboring words. As fastText approach was proposed using skip-gram model with embedding dimension of 300 in paper \cite{fastText_mikolov} and skip-gram model has been proved to be effective compared to CBOW model in the existing literature \cite{word2vec_skipgram_mikolov}. Besides, comparing different embedding dimensions, the existing study has shown that a dimension of 300 achieves both high accuracy and reasonable training time \cite{api_embedding, word2vec_skipgram_mikolov}. Since the average API description length in our dataset is 11.15, we considered the context window size 5. This context window size is also the default value in Gensim package and is used in the existing literature (e.g., \cite{deeptip}). For building the word dictionary, we kept the words that appear at least once in the training set (min\_count) so that the dictionary does not miss any word from the training set. We used word\_ngram to enrich the word vectors with subword (n-grams) information, which is required to be 1 for fastText implementation in Gensim. For other parameters, we have considered the default settings in Gensim, which is a common practice in the existing literature \cite{fastText_mikolov, deeptip}.\\


\noindent\fbox{
    \parbox{0.97\linewidth}{%
        \textbf{Hyper-parameter settings of Word Embedding:} context window=5, embedding dimension=300, min\_count=1, workers=3, model=skip-gram, hierarchical softmax(hs)=1, word\_ngrams=1
    }
}
\newline

We implemented the CNN model using Keras \cite{keras} library. The CNN model was tuned with various hyper-parameters like number of hidden nodes, batch size, and number of filters using grid search. We used Grid search technique for model hyper-parameter optimization, which is used in the existing deep learning-based techniques (e.g., \cite{cnn_sentence}). Grid search takes every combination of a list of values of the hyper-parameters and chooses the best combination based on the cross-validation score. Besides, grid search can be parallelized to reduce the hyper-parameter optimization time. Thus, grid search is widely used in the existing literature \cite{cnn_sentence, sec_orch_caise} for hyper-parameter optimization. For conciseness, we only reported the best parameter setting. We set the filter window size = [3, 4, 5], which denotes that the convolutional filters slide over 3, 4, and 5 words, respectively. We used 100 filters for each filter size. 


We used early stopping \cite{cnn_sentence} to stop the training epoch once validation accuracy started to decrease with patience=50. We added dropout and L2 regularization (L2R) \cite{cnn_sentence} to the network to mitigate overfitting. We randomly shuffled the dataset and splitted the data as follows: 80\% for training, 20\% for testing, and we performed 10-fold cross-validation. The number of cross-validation repeats that we performed was 10. These splitting and repeat scheme are commonly used in the existing deep learning-based methods (e.g., \cite{split_scheme, deeptip}). Our corpus includes high number of API descriptions in different writing forms containing rich contextual information regarding the APIs (described in Section \ref{subsec: rq3}). This divergence makes our corpus suitable to train our neural architecture and test it's performance. The experimental results present that our training corpus is sufficient for training a neural network with high performance.\\

\noindent\fbox{%
    \parbox{0.97\linewidth}{%
        \textbf{Hyper-parameter settings of CNN:} batch=64, epoch= early stopping criteria with patience=50, dropout=0.5, embedding dimension=300, filter sizes=(3, 4, 5), number of hidden node=100, L2R=0.0001, number of filters=100
    }%
}

\subsection{Baseline Approach}\label{subsec:baseline_approach}
We used the state-of-the-art W2V-IDF similarity score based approach as the baseline approach, which was also adopted for the research on Java API recommendation \cite{from_word_to_doc, biker, biker_tool}. 
To develop W2V-IDF approach we built a word embedding model \cite{word2vec_mikolov} and word IDF (Inverse Document Frequency) vocabulary from the pre-processed security tool API description from API documentation. Here, IDF of a word denotes the inverse of total number of API descriptions that have the word. To recommend APIs, similarity scores among the query and API descriptions were calculated. The similarity score was calculated based on IDF-weighted similarity on top of cosine similarity of the word embedding following similar approach as used in \cite{from_word_to_doc, biker}.   



\subsection{Evaluation Metrics}\label{subsec:eval_metrics}
APIRO and the baseline approach were evaluated using Top-K Accuracy (Top-K acc), Mean Reciprocal Rank@K (MRR@K), and Mean Precision@K (MP@K). These evaluation metrics have been commonly used for information retrieval and recommendation systems in the existing works \cite{from_word_to_doc, rack, polisis}.

\textbf{Top-K Accuracy (Top-K acc)} is the percentage of queries that a recommendation system that can recommend the correct API among the Top-K results.

\begin{equation}
    Top-K Acc (Q) =  {\frac{\sum_{q \in Q} Correct\_API(q,Top-K)}{|Q|}}\%
\end{equation}

Here, Q represents the set of all queries in test data. Correct\_API(q, Top–K) returns  1 if a correct API exists in the Top-K results and 0 otherwise.

\textbf{Mean Reciprocal Rank@K (MRR@K)} is useful when we want our system to return the best relevant item at a higher rank. Reciprocal rank@K is the multiplicative inverse value of the correct API's rank in a recommendation system's Top-K results. MRR@K is the average of RR@K for all the queries in test data.

\begin{equation}
    MRR@K (Q) = {\frac{1}{|Q|}}\sum_{q \in Q} {\frac{1}{Rank_q}}
\end{equation}
Here, Q represents the set of all queries in test data and $Rank_q$ represents the rank of the correct API for a query.

\textbf{Mean Precision@K (MP@K)} is the mean value of precision@K for all the queries in test data. Precision@K is the ratio of True\_Positives@K (TP@K) and the summation of TP@K and False\_Positives@K (FP@K).

\begin{equation}
    MP@K (Q) = {\frac{1}{|Q|}} \frac{TP@K}{(TP@K)+(FP@K)}
\end{equation}

\textbf{Performance Gain} is used to determine whether APIRO is better than the baseline approach. The percentage of the performance difference between two approach are calculated to measure performance gain using equation \ref{equ:PG}.

\begin{equation}\label{equ:PG}
    \%PerfDiff = {\frac{(Perf_{APIRO})- (Perf_{baseline})}{Perf_{baseline}}}*100
\end{equation}

A positive value of the performance gain indicates that the performance of APIRO is better than the baseline approach, while a negative value indicates that the performance of APIRO is poor than the baseline approach.
\section{Experimental Results}\label{subsec:exp_res}
This section presents the experiment results for answering our three proposed research questions.
\subsection{RQ1. Applicability of Data Augmentation Techniques } \label{subsec: rq1}
\textbf{Motivation:}
APIRO leverages diverse text augmentation techniques to mitigate the data availability constraint and enrich semantic variation in the security tool API corpus.  
It is challenging to find suitable approach for text data augmentation as all the data augmentation techniques do not always preserve the context and semantics. For instance, a pre-processed data from API documentation `specifies maximum number flowbit tag use within rule set' is augmented using WordEmbsAug ($Aug_6$) approach with data source `\textit{GloVe Common Crawl (action = insert,  dimension = 300)}'. The output of this augmentation technique is `sicheianytimes.com specifies maximum number 1tablespoon flowbit tag use within rule set'. 
In augmented data, the tokens `\textit{sicheianytimes.com}' and `\textit{1tablespoon}' are semantically irrelevant and does not preserve the context of the original sentence. In order to answer RQ1, we investigated and identified various text augmentation techniques which retain the semantic similarity and relatedness of the sentences when applying them for security tool API data. 

\textbf{Approach:} 
To answer RQ1, we analyzed a data sample of size 555, which included 36 augmented data and original data of each of the 15 APIs from three security tools as described in Section \ref{subsec:methodology}. We labeled selection value to all the data of the data sample. 19 out of 36 augmentation techniques were selected, which had selection score ($S_{score}(DAT_\alpha)$) (calculated using Equation \ref{eq:select}) greater than the mean selection score, i.e., $M_{score}$. Here, $M_{score}$=53.5, which is the mean value of the selection scores of 36 augmentation techniques.

\begin{figure*}[h]
  \centering
  \includegraphics[scale = 0.35]{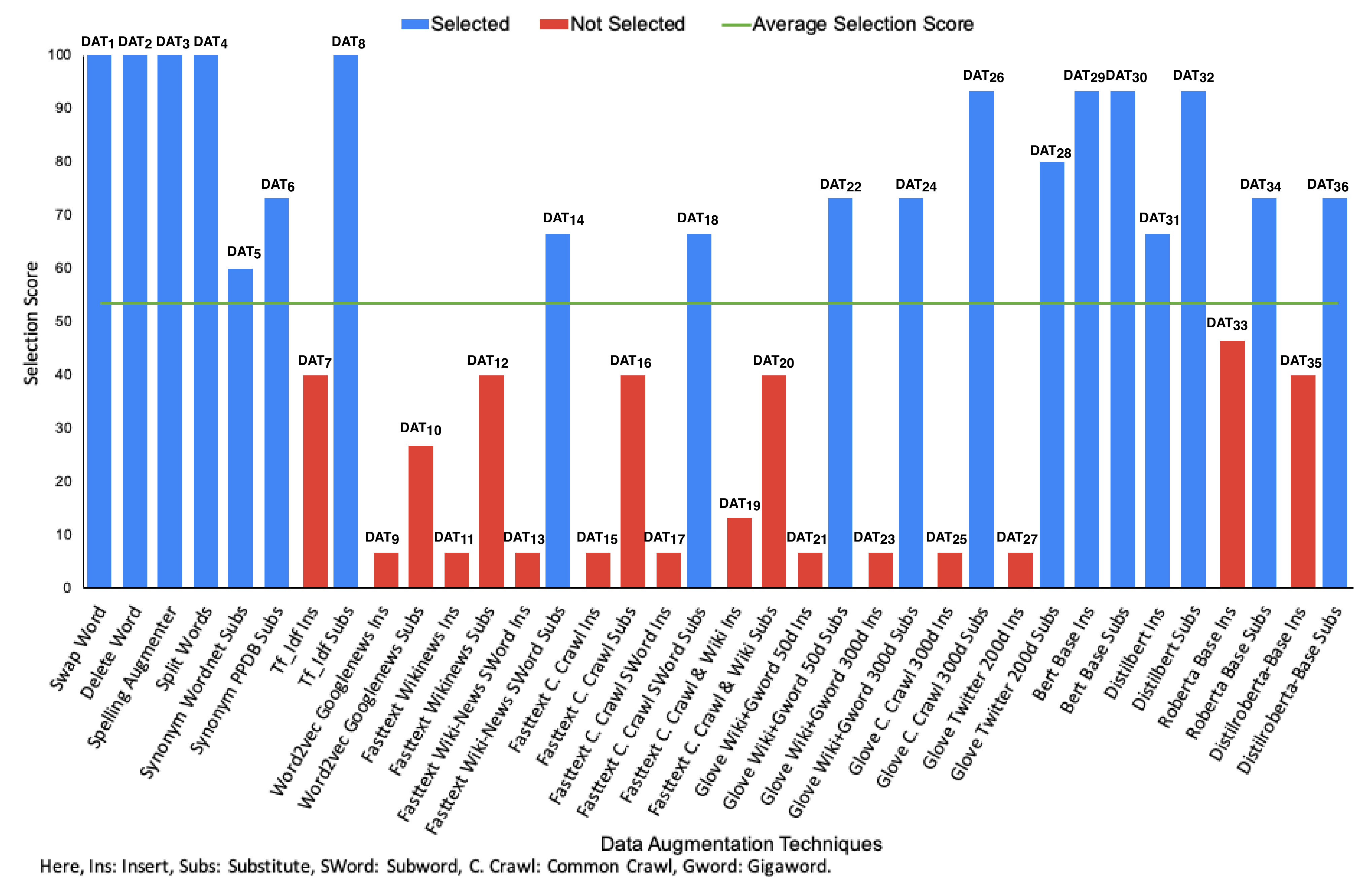}
  \caption{Data augmentation techniques selection based on selection score}
  \label{fig:dataaug_select}
   \vspace{-10pt}
\end{figure*}

\textbf{Results:}
Fig. \ref{fig:dataaug_select} demonstrates the graphical analysis of data augmentation techniques ($DAT_i$), where augmentation techniques were chosen based on the selection score. Here, i= 1, 2,..., 36 and the augmentation approaches ($Aug$) were described in Section \ref{subsubsec:data_aug}. Our analysis shows that the selection score for $Aug_1$(i.e., \textit{swap word}, \textit{delete word}), $Aug_2$ (i.e., \textit{spelling Augmenter}), and $Aug_3$ (i.e., \textit{split words}) techniques covering ($DAT_{1-4}$) are very high and got selected. The reason behind the high selection score is that light textual noise and spelling error injection to the security API corpus do not change much of the semantic similarity and context. The selection scores of $DAT$ under $Aug_4$ (\textit{synonym WordNet} (i.e., $DAT_5$) and \textit{PPDB} (i.e., $DAT_6$)) are slightly above $M_{score}$. The reason is that replacing synonym may not always give aligned semantic meaning to the original data (e.g., command line switch is replaced by statement line switch by WordNet). Our experimental results for $DAT_{1-6}$ (i.e., shown in Fig. \ref{fig:dataaug_select}), reflect the findings of Claude et al. \cite{nlp_cloud}. 

$Aug_5$ includes \textit{TF-IDF insertion (i.e., $DAT_7$)} and \textit{TF-IDF substitue (i.e., $DAT_8$)}. Injecting new word to random position according to TF-IDF statistics of words (i.e., $DAT_7$) on sample security tool API corpus provided inconsistent word. For instance, $DAT_7$ added the tool name `snort' in API description of other tools such as Limacharlie and MISP. Hence, $DAT_7$ had selection score less than $M_{score}$ but $DAT_8$ had selection score greater than $M_{score}$ as it does not create such kind of inconsistency. 

For $Aug_6$ (i.e., WordEmbsAug), we performed `insert' and `substitute' action for augmenting text using similar word based on different word embedding techniques which are Word2vecAug, fastTextAug, and GloVeAug. Fig. \ref{fig:dataaug_select} shows that in {$DAT_{9-28}$} we leveraged different pre-trained vectors of these word embedding techniques such as Word2vecAug (e.g., Googlenews), fastTextAug (e.g., Wikinews, Common crawl), and GloVeAug (e.g., Wiki+Gigaword, Twitter). Our analysis shows that for $Aug_6$ the selection score of $DAT$ with `insert' action were lower than $M_{score}$. These $DAT$ help to inject noise to the API corpus by picking word from vocabulary randomly, which justifies the lower selection score for $DAT_{(9,11,13,15,17,19,21,23,25,27)}$. 

For $Aug_6$ with `substitute' action, the selection scores were less than $M_{score}$ for four pre-trained embedding vectors, which were Word2vec Googlenews ($DAT_{10}$), fastText Wikinews ($DAT_{12}$), fastText Common Crawl ($DAT_{16}$), and fastText Common Crawl \& Wiki ($DAT_{20}$). Using these pre-trained vectors provided words that were not semantically similar or relevant to the security API data. However, `substitute' action for both fastText Wikinews with sub-word ($DAT_{14}$), fastText Common Crawl with sub-word ($DAT_{18}$) had selection score greater than $M_{score}$. The reason behind this can be the usage of sub-word information by these two $DAT$, which helps generate better similar words for rare and OOV words in the dataset. Our analysis resulted in selection of the GloVe embeddings with `substitute' action that are wiki2014+gigaword 50d ($DAT_{22}$), wiki2014+gigaword 300d ($DAT_{24}$), Common Crawl 300d ($DAT_{26}$), and Twitter 200d ($DAT_{28}$) with high selection score. The high selection score can be justified as GloVe does not rely just on local context information of words like in word2vec, but incorporates global (over the whole corpus) word co-occurrence statistics to generate similar word vectors. 
Fig. \ref{fig:dataaug_select} shows the higher selection score of $DAT_{29-32,34,36}$ of contextualWordEmbsAug ($Aug_7$) with data source Bert Base, Distilbert Base for both `insert' and `substitute' and Roberta Base and Distilroberta Base for `substitute' action. Unlike WordEmbsAug (e.g., Word2vec), BERT language model predicts a word for insertion rather than picking word randomly, which justifies the good selection score for `insert'. Besides, substitution uses surrounding words as a feature to predict the target word. Hence, compared to WordEmbsAug approach, the generated text in $Aug_7$ are more semantically coherent as the model takes context into account while making predictions. Overall, our qualitative analysis discarded 47.2\% augmentation techniques (17 out of 36 $DAT$), which did not provide semantically relevant API data. Our investigation of different augmentation techniques reveals that all the augmentation techniques are not suitable. Rather, the augmentation techniques that tend to preserve the semantic and context are more suitable. 

\subsection{RQ2. Performance Comparison}\label{subsec: rq2}
\textbf{Motivation:} The goal of APIRO is to automatically recommend appropriate security tool APIs for natural language queries. Ye et al. \cite{from_word_to_doc} and Huang et al. \cite{biker} used W2V-IDF based approach to recommend Java API to query. Huang et al. \cite{biker} considered SO as a datasource and presented both SO only and documentation only (i.e., here referred as W2V-IDF) recommendation evaluation results. There is a lack of security tool API related questions in SO (described in Section 1). So, our proposed framework APIRO considers API documentation from official repositories which is considered as reliable, trustworthy, and updated data source \cite{api_doc_survey}. Thus, RQ2 evaluates the effectiveness of APIRO and how it compares with the baseline W2V-IDF approach. Further, we evaluate the impact of clustering, embedding choice, immutable word, and data augmentation in the performance of APIRO.

\textbf{Approach:} To answer RQ2, APIRO was compared with the developed baseline approach W2V-IDF (described in Section \ref{subsec:baseline_approach}). For a fair and valid comparison, the approaches were run under the same environment and evaluated on the same test data representing diverse types of queries. The evaluation test data was created from `post-processed augmented corpus' as described in Section \ref{subsec:methodology}. Ye et al. \cite{from_word_to_doc} and Huang et al. \cite{biker} did not consider any augmentation approach, the train data for W2V-IDF is the pre-processed and clustered data from original API documentation as described in Section \ref{subsec:methodology}. 
We investigated Top-K Acc, MRR, MP@K, train-test time to compare the performance of APIRO with baseline approach W2V-IDF. To show the impact of different factors in APIRO, we present the corresponding approaches as follows:

i. \textbf{Impact of clustering:} To show the impact of clustering in APIRO, we compared the performance of APIRO with the \textit{\textbf{APIRO: without clustering}}' variant, where API clustering (based on same API description) is not performed.  

ii. \textbf{Impact of embedding choice:} To analyse the impact of non-domain-specific pre-trained embedding, we used the general-purpose word embedding `glove.6B.300d.txt', which is GloVe \cite{glove} pre-trained embedding on Wikipedia data, where embedding dimension is 300. Besides, we performed the comparison of fastText embedding with word2vec embedding choice.

iii. \textbf{Impact of immutable words}: We showed the impact of immutable words on the augmented data by performing manual validation on sample augmented data without considering immutable words for the selected \textit{DATs} only. The details of manual validation is discussed in section \ref{subsubsec:data_aug} and section \ref{subsec:methodology}. The Cohen’s Kappa value for the manual validation to check whether augmentation without immutable words changes the semantics and context of an API description is 0.74 and there is 91.3\% agreement, that indicates substantial agreement. To show the impact of immutable words in APIRO, we compared the performance of APIRO with the `\textit{\textbf{APIRO: without immutable words}}' variant, where immutable words were susceptible to augmentation.

iv. \textbf{Impact of different \textit{DAT}}: To analyse the effect of individual \textit{DAT} on the performance of APIRO, we compared APIRO with various variants of it. Each variant was created by removing one of the selected \textit{DATs} from the training set of APIRO. We then checked which \textit{DAT} had the most important role in the performance of APIRO.





\textbf{v. Impact of learning model}: To analyse the impact of RNN as a learning model for security tool API data, we implemented the commonly used variant of RNN \cite{textshield, vuldeepecker}, which is Bidirectional Long Short Term Memory (Bi-LSTM) \cite{Bi_LSTM}. While RNNs can suffer from ineffective model training due to Vanishing Gradient (VG) problem \cite{vg_prob}, VG problem can be addressed by using Long Short-Term Memory (LSTM) cell \cite{Bi_LSTM, lstm}. Bi-LSTM exploits both forward and backward direction on the input sequence to get text representation. The used Bi-LSTM structure includes Bi-LSTM layer, dense layer, and a softmax layer. We use the same security-tool specific fastText embedding for CNN and RNN implementation. Our BiLSTM is designed with one bidirectional layer of 128 hidden units, which is a common practice for textual data \cite{textshield}. Both CNN and RNN are trained with Adam optimizer. The optimal hyper-parameters such as maximum training epoch, dropout rate, batch size are tuned separately for the RNN model. A dropout \cite{lstm} and recurrent dropout rate of 0.2 are used and the batch size is set to 64. Besides, we used early stopping \cite{early_stop}, where the training is stopped if the performance of the model did not improve for 50 epochs.


\begin{figure}[h]
  \centering
  \includegraphics[width=\textwidth]{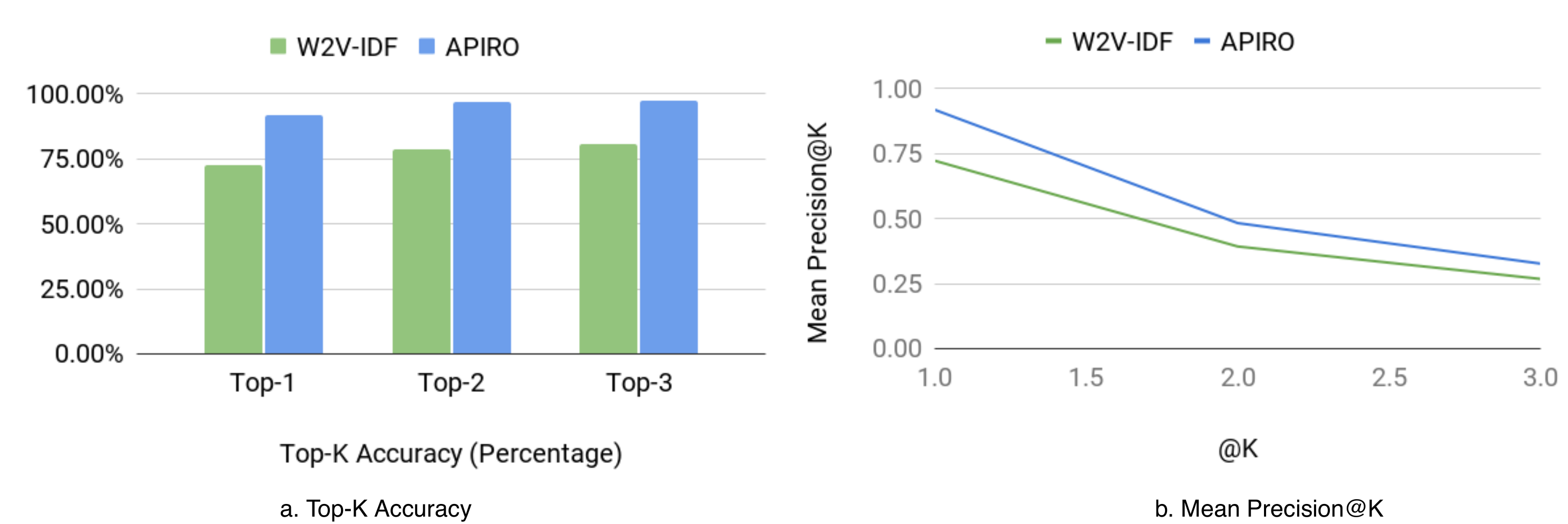}
  \caption{(a) Top-K Accuracy and (b) Mean Precision@K of APIRO and W2V-IDF for security tool API recommendation}
  \label{fig:rq3_comp}
  \vspace{-10pt}
\end{figure}

\textbf{Results:} We evaluate both the effectiveness and efficiency to compare APIRO with the baseline.

\textit{Effectiveness of APIRO:} Fig. \ref{fig:rq3_comp}(a) shows the comparison of Top-K Acc of APIRO and W2V-IDF approach for $k$=1, 2, 3. As shown in Fig. \ref{fig:rq3_comp}, Top-1 Acc of APIRO is 91.9\%. This result demonstrates that our approach can recommend a relevant API correctly with 91.9\% accuracy. On the other hand, W2V-IDF approach has 72.4\% Top-1 Acc, where APIRO is 26.9\% more efficient than W2V-IDF. APIRO is 23.0\% and 20.9\% improved compared to W2V-IDF in terms of Top-2 and Top-3 Acc, respectively. It is desired to gain a high Top-K score for low values of K so that a user requires to process less information before reaching correct API. The promising Top-3 Acc of APIRO suggests that it can help user to find relevant API with less time and effort as a user has to process less number of APIs to have the desired relevant API.
Fig. \ref{fig:rq3_comp}(b) presents the MP@K for APIRO and W2V-IDF for different values of K. Higher values of K is less desired as it represents that more APIs need to be presented to users for relevant API, thus providing low value of MP@K. For MP@k, APIRO achieves value of 0.92, 0.48, and 0.33, which outperforms W2V-IDF as W2V-IDF gains 0.72, 0.39, and 0.27 for k=1, 2, and 3, respectively. 

Table \ref{tab:mrr_comp} shows the MRR of APIRO and W2V-IDF. MRR of APIRO is 0.94 with performance gain of 23.7\% compared to W2V-IDF. The high MRR value of 0.94 for APIRO denoting that the best relevant API is at a higher-ranked position among the top-3 recommended API. Hence, APIRO notably outperforms the developed state-of-the-art baseline W2V-IDF in terms of Top-K Acc, MRR, and MP@K, respectively.

\begin{table}[]
\centering
\caption{Mean reciprocal rank of APIRO and W2V-IDF for security tool API recommendation}
\label{tab:mrr_comp}

\begin{tabular}{lc}
\toprule
Approach & Mean Reciprocal Rank (MRR)\\ 
\midrule
Baseline: W2V-IDF & 0.76\\
Proposed: APIRO & 0.94\\
Improvement & 23.7\%\\
\bottomrule
\end{tabular}
 \vspace{-10pt}
\end{table}



\textit{Efficiency of APIRO:} APIRO requires 28.7 minutes of training time to perform the data augmentations, build word embedding corpus (i.e., fastText) and train the learning model (CNN). About half an hour of training time is reported to be acceptable by the existing study \cite{intention_mine1}. During the initial training of APIRO, if the organization considers all the security tools used in their SOC, then frequent re-training will not be required. However, the API recommendation support for security incident response will not be interrupted, as the re-training can be done independently and the existing model can be replaced by the re-trained model accordingly. Thus, re-training can be done offline and would not cause actual down-time of the recommendation service. Further, GPU can be used to improve APIRO time-efficiency (i.e., accelerate the CNN training time by GPU \cite{cnn_gpu_1, cnn_gpu2}). W2V-IDF gains faster training time (1.8 sec), which creates word embedding corpus and IDF dictionary. However, W2V-IDF can not achieve a high accuracy compared to APIRO.

APIRO only requires 2.9 seconds to recommend APIs for thousands of queries (2608 queries). In contrast, W2V-IDF needs 39.6 minutes to recommend APIs for the 2608 queries, as for each query W2V-IDF needs to calculate the similarity score of the query to each of the API descriptions available in the input data. Delay in API recommendation for security incident response may cause delay in the execution of the incident response action. This delay in incident response execution in SOC can cause huge loss to the organization due to the adverse effect of a security attack. Thus, the faster prediction time of APIRO denotes that APIRO is efficient for the use of security tool API recommendation in practice for the time-critical SOC environment.



\begin{figure}[h]
  \centering
  \includegraphics[scale=0.4]{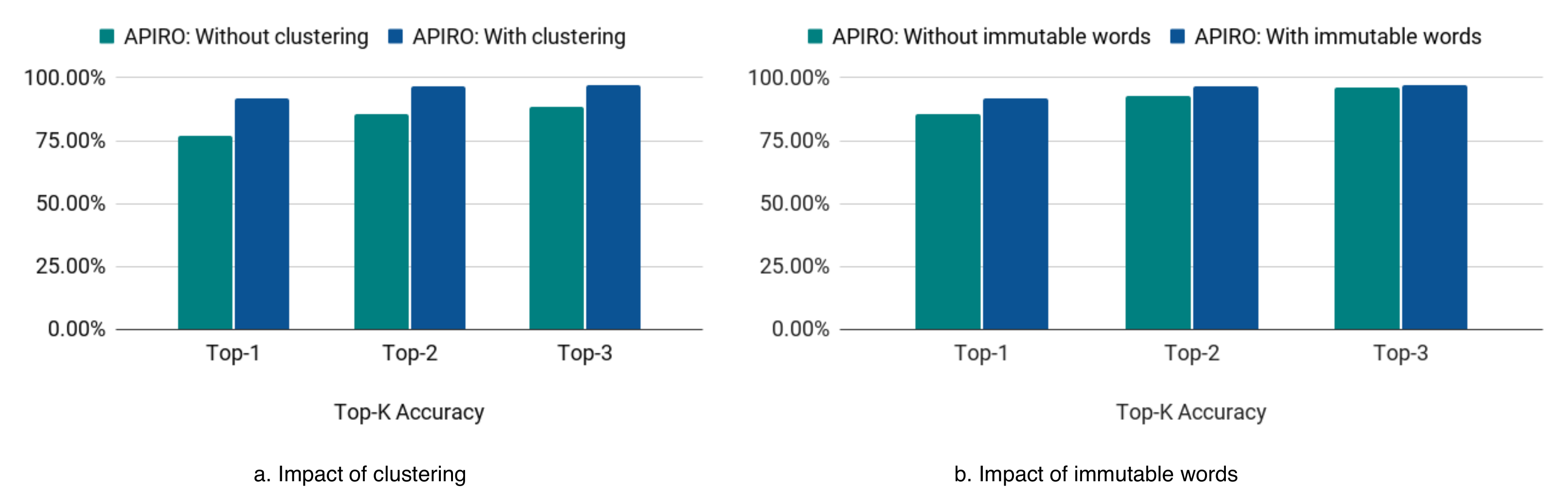}
  \caption{(a) Impact of clustering (b) Impact of immutable words in APIRO for security tool API recommendation}
  \label{fig:clustering}
   \vspace{-10pt}
\end{figure}

\textbf{Impact of factors on APIRO:} The impact of different factors in APIRO are presented as follows: 

\textbf{i. Impact of clustering:} Fig. \ref{fig:clustering}(a) shows that APIRO is 19.2\%, 13.2\%, and 10.2\% improved compared to `APIRO without clustering' in terms of top-1, top-2, and top-3 accuracy, respectively. Without clustering, having multiple APIs with identical description is challenging for the model to decide which API to recommend. Besides, in our corpus, the maximum number of APIs in a cluster is 9 and APRIO recommends relevant top-3 APIs. Without clustering, even if APIRO recommends the top-3 relevant APIs, still some of the APIs are missing in the recommended list even though the missing APIs have the exact same description. By clustering the APIs having the same description, APIRO is able to recommend the full list of APIs with the same API description as a cluster in its top-3 recommended result justifying the benefit of using clustering in APIRO. 


\textbf{ii. Impact of embedding choice:} Use of domain-specific fastText embedding achieved 1.7\%, 1.3\%, and 1.1\% improvement compared to the existing pre-trained Glove embedding in APIRO in terms of top-1, top-2, and top-3 accuracy, respectively. Since the existing pre-trained embedding is trained on general purpose (e.g., Wikipedia, Google-news) finite dataset, embeddings are not available for infrequent words. Hence, domain-specific (i.e., security tool API) words are not always supported and presented as zero matrix by the existing pre-trained embeddings. 
The pre-trained Glove dataset only covers 80.3\% of the words from our training dataset and the remaining 19.7\% of words are presented as zero matrix. For example, the security tool-specific words such as ‘Limacharlie’, ‘pcap’, and ‘MISP’ do not have any embedding in the existing pre-trained Glove embedding. Thus, the security tool API-specific embedding model provided better prediction accuracy compared to the general-purpose pre-trained embeddings, which aligns with the findings of the existing research \cite{polisis, concept_drift}, where domain-specific embedding proved to perform better.

Use of fastText embedding achieved 1.2\%, 1.1\%, and 0.7\% improvement compared to word2vec embedding in APIRO in terms of top-1, top-2, and top-3 accuracy, respectively. Word2vec cannot generate embeddings for OOV words, whereas fastText generate the word vectors for OOV words by averaging the vector representation of its n-grams \cite{fastText_mikolov}. In our test dataset, 12.4\% of the words were OOV words. 
For example, in our dataset for the OOV words `listify’ and `datagrams’, word2vec fails to generate an embedding as the words are not available in the vocabulary. However, fastText is able to create embedding for these words and the top-most similar words for `listify’ and `datagrams’ by fastText are `listing’ and ‘datagram’, respectively. FastText provided better prediction accuracy compared to word2vec embeddings, which aligns with the existing research \cite{fastText_mikolov, polisis}, where the OOV issue is resolved using fastText embedding. Hence, for better generalization of APIRO to answer free-form query fastText is a preferred choice to encounter OOV words.

\textbf{iii. Impact of immutable words}: We compared the original API descriptions with the augmented API descriptions that were augmented without considering immutable words. Our analysis found that 21.7\% augmented API descriptions did not hold semantic similarity due to performing augmentation without considering immutable words. For example, while using \textit{Wordnet-based synonym replacement} augmentation technique, the word `PID' is replaced with `pelvic inflammatory disease' in corresponding augmented API description, which (i.e., PID) is part of our immutable word list. Here, `pelvic inflammatory disease' is changing the original context, so the corresponding augmented API description is considered not semantically relevant to the original API description. Similarly, the immutable words `log' (i.e., log file) is augmented with the word `logarithm' and `galaxy' (i.e., MISP galaxy is a large cluster that can be attached to MISP events or attributes) is augmented with the word `mega star'. Both `logarithm' and `mega star' are changing the original context and are not semantically relevant to the corresponding security tool API. Providing semantically irrelevant data is detrimental to the learning model. `Garbage in, garbage out' is a well-established concept among ML/DL community \cite{garbage2}. Various existing literature \cite{garbage1, garbage2} have supported inconsistent training data can heavily impact the learning model's performance. Thus, the use immutable word mitigates the issues of having semantically irrelevant words in the learning model.
Fig. \ref{fig:clustering}(b) further shows that APIRO is 7.1\%, 3.9\%, and 1.5\% improved compared to the `APIRO without immutable words' variant in terms of top-1, top-2, and top-3 accuracy, respectively. Hence, the result justifies the significance of immutable words in APIRO.

\begin{figure}[h]
  \centering
  \includegraphics[scale=0.4]{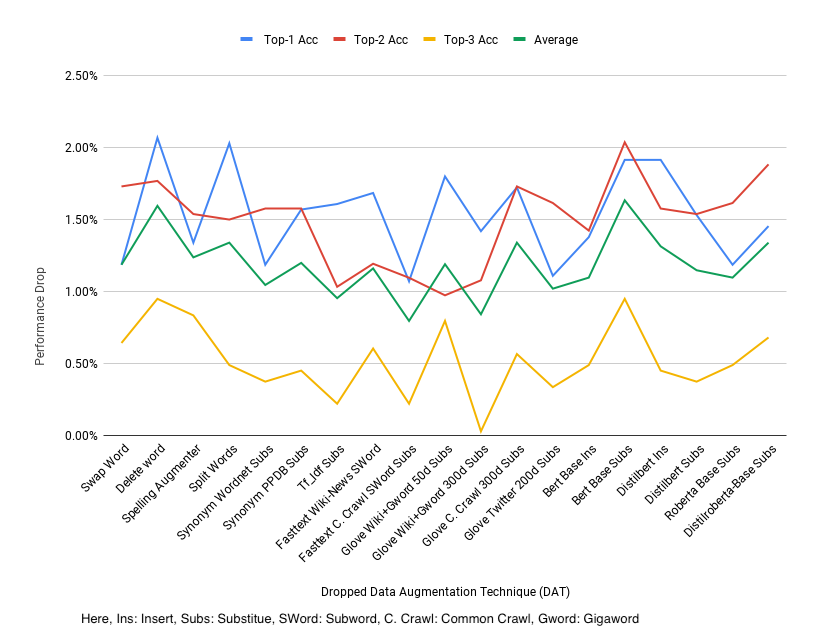}
  \caption{Effect of dropping different data augmentation techniques in APIRO for security tool API recommendation}
  \label{fig:rq2_aug_drop}
   \vspace{-10pt}
\end{figure}

\textbf{iv. Impact of different \textit{DAT}}: Fig \ref{fig:rq2_aug_drop} shows the impact of data augmentation in terms of the top-K accuracy of each variant. It shows low performance after dropping a \textit{DAT} training data compared to while considering all \textit{DAT}. The results show that no matter which \textit{DAT} we dropped, it hurts the performance of APIRO. This verifies the importance and effectiveness of the selected \textit{DATs}. The top three variants with the maximum performance drop are the variants that removed the following \textit{DATs}: `Bert Base Subs' (i.e, $DAT_{30}$), `Delete Word' (i.e., $DAT_2$), and `Distilroberta-Base Subs' (i.e., $DAT_{36}$). Here, $DAT_{30}$ and $DAT_{36}$ are contextual word embedding-based augmentation techniques that substitute contextually similar words to augment the data. The three variants with the minimum performance drop are the variants that removed \textit{DATs}: `Fasttext C. Crawl SWord Subs' (i.e, $DAT_{18}$), `Glove Wiki+Gword 300d Subs' (i.e., $DAT_{24}$), and `Tf\_Idf Subs' (i.e., $DAT_8$). Here, $DAT_{18}$ and $DAT_{24}$ are non-contextual word embedding-based augmentation techniques. Hence, the contextual word embedding-based augmentations show more impact on APIRO performance. This aligns with our earlier observation in RQ1, where 6 out of 8 (i.e, $DAT_{29}-DAT_{36}$) contextual word embedding-based augmentation techniques got selected, whereas only 6 out of 20 ($DAT_{9}-DAT_{28}$) non-contextual augmentation techniques got selected. 
The results show that dropping any of the $DATs$ caused performance drop to APIRO. Thus, the different $DATs$ that we applied complements each other to provide better learning experience to the APIRO model.





\begin{figure}[ht]
  \centering
  \includegraphics[scale=0.4]{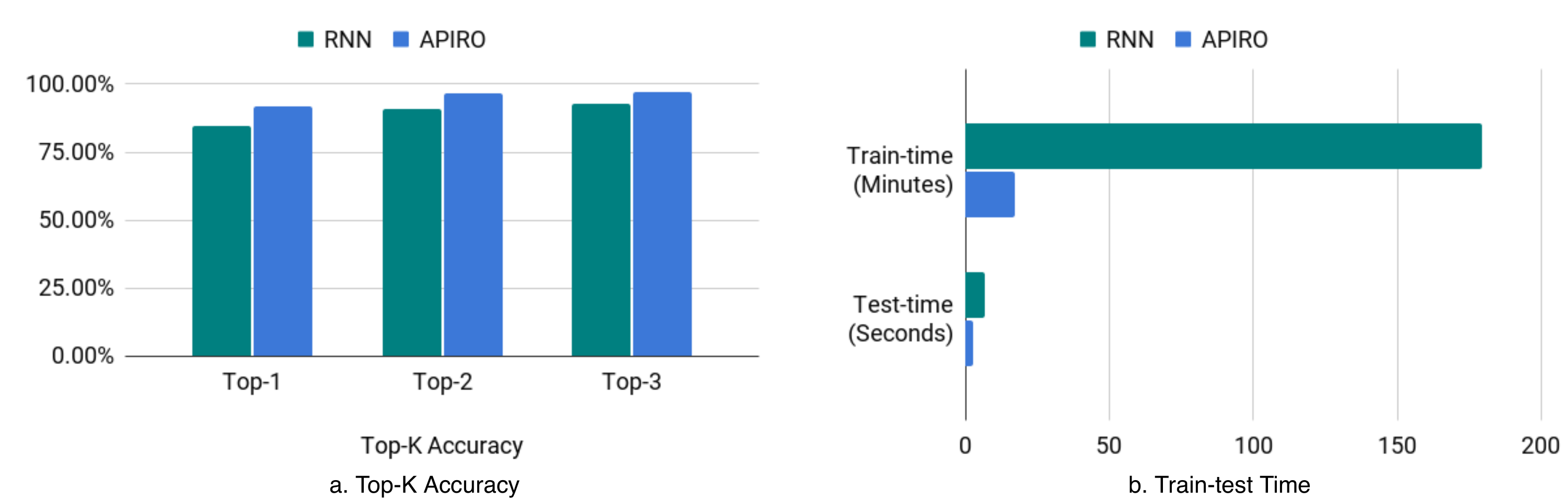}
  \caption{(a) Top-K accuracy and (b) Train-test time of RNN and APIRO for security tool API recommendation}
  \label{fig:blstm}
  \vspace{-10pt}
\end{figure}

\textbf{v. Impact of learning model:} Fig \ref{fig:blstm} shows the top-k accuracy and train-test time comparison of RNN and CNN (in APIRO). Our implementation evaluation aligns with the observation of the existing literature \cite{rnn_over_cnn1, rnn_over_cnn2}, as CNN outperformed RNN by 6.6\% on average in terms of top-k accuracy. Moreover, to achieve this accuracy RNN required 179.6 minutes training time. RNN required 10.5 times more training time and 2.4 times more testing time compared to CNN. It further shows that for security tool API data, CNN is both effective and efficient over RNN.


\subsection{RQ3. Effectiveness of Answering Free-form Natural Language Query}\label{subsec: rq3}
\textbf{Motivation:}
The goal of this RQ3 is to investigate the robustness of the APIRO to answer different categories of free-form natural language queries for recommending security tool API. SOC team members may not always be able to describe the API related query clearly. Query structure can vary from user to user, depending on users' expertise level, experience, and communication skill. For example, among 8.5 million SO queries, 2.1 millions (24.7\%) of them had issues like spelling, grammar, or formatting which are very frequent issues of queries in SO \cite{query_type_reform_mistake}. Besides, to avoid having a few or empty search results, some users may deliberately limit a query’s length which results in a short query \cite{query_length}.
For example, the mean and 75th percentile of the length of the queries (i.e., the total number of words) on QA sites like Stack Overflow are 3.6 and 4, respectively \cite{query_reform_so}. Moreover, different words (e.g., synonyms, contextually similar words) can be used to convey the same concept, which is a common challenge while answering queries \cite{rack}. Study \cite{query_reform_so} showed that 19.1\% of their analyzed SO queries had spelling and syntax issues, while 15.1\% of the queries had detailed or unnecessary words. All these characteristics of queries make answering free-form queries challenging. Hence, we identified five major challenging free-form query categories for evaluating the robustness of APIRO model across these query categories and compare the performance with the baseline. The evaluation query categories are - \textit{Q1. erroneous query, Q2. query including synonyms, paraphrases, Q3. query including contextually similar words, Q4. unordered query, and Q5. short query}. Table \ref{tab:q_cat_example} provides  examples of each query category and example APIs.



\begin{table}[ht]
\caption{Free-form natural language query categorization and test query generation}
  \label{tab:q_cat}
\begin{tabularx}{\textwidth}{p{5.91cm}p{8.8cm}}
 \hline
\textbf{Query category} & \textbf{Test query (included $DAT_i$)} \\
 \hline
Q1. Erroneous query & $DAT_3, DAT_4$ \\ 
Q2. Query including synonyms, paraphrases & $DAT_5, DAT_6$\\
Q3. Query including contextually similar words & $DAT_8$, $DAT_{14}$, $DAT_{18}$, $DAT_{22}$, $DAT_{24}$, $DAT_{26}$, $DAT_{28-32}$, $DAT_{34}$, $DAT_{36}$\\
Q4. Unordered query & $DAT_1$ \\
Q5. Short query & $DAT_2$\\
\hline
\end{tabularx}
 \vspace{-8pt}
\end{table}
\textbf{Approach:} 
We evaluated how APIRO and W2V-IDF perform to answer the five identified challenging query categories in terms of Top-K Accuracy. We selected the test data set from the Augmented corpus which provides meaningful queries.
As discussed in section \ref{subsec:methodology}, we only considered the realistic queries generated from the selected data augmentation techniques that preserves semantic similarity or relevance. We divided the `processed augmented corpus' into 20 groups which included 19 selected augmented data groups and one original data group. The 19 augmented data groups were generated using the 19 selected augmentation techniques (selection details discussed in RQ1). We used one augmented data group for testing, while we used all the other 19 out of 20 data groups for training. This process was repeated with every augmented data group as test data while keeping other data groups as training data. Each time, the test data was kept unseen to the model while training. Next, each of the 19 augmented test data group was mapped to the five query categories. Table \ref{tab:q_cat} shows the mapping of the test data groups to the query categories. To calculate the Top-K Acc of a query category, which was mapped to multiple test data groups, we calculated the average Top-K Acc of the corresponding test data groups. The same 19 test data groups were used to evaluate W2V-IDF approach for comparing its' performance with APIRO for different query categories.\\ 


\textbf{Results:} Table \ref{tab:q_cat_example} presents examples of each free-form natural language query category with corresponding API and tool. Query incorporating `how' phrases (e.g., how to, how do I, how can I) followed by task description are usually the most frequent query pattern in question answering forums \cite{query_reform_so}. The prepossessing steps of APIRO remove the stops words such as \textit{(how to, how do I, how can I)} and punctuation such as '?' (details in section \ref{subsec:pre-processing}. 
Therefore, the pre-processed query have the task description which can be queried in varied ways by SOC teams. By \textit{Q1}, we refer to the queries that can have spelling or typing mistakes such as `commmunity’ instead of `community’. \textit{Q2} refers to queries that include synonyms or paraphrases representing the same concept. For example, the synonym `lookup’ is used to get the API for performing a `search'. By \textit{Q3}, we refer to queries that do not contain exact synonym or paraphrase in API description rather include words that are contextually similar (i.e., words that frequently appear in the same context). The example query \textit{Q3} includes the term `intrusion’, which frequently appears with the term `detection’ in general. Here, the aim is to find an API for Limacharlie to delete detection and response rule and the API documentation of Limacharlie does not contain the word intrusion. By \textit{Q4}, we refer to a query where the order of words can be grammatically incorrect. \textit{Q4}) has the keywords (e.g., payload, available) but the ordering of the words may vary (e.g., `list the available payloads', `how to listify payloads available'). \textit{Q5} refers to short queries, which may miss some important key terms. The average length of the short query in our test data is 4.9, whereas the average length of queries in the rest of the dataset is 7.4. For example, to remove a yara rule source, a user query is `How to remove yara rule?'. In this example short query (i.e., \textit{Q5}), the keyword `source' is missing. The missing keyword `source' is important to clearly distinguish the API (limacharlie.Replicants.Yara.removeSource) with another API that removes the rule itself (limacharlie.Replicants.Yara.removeRule).
\begin{table}[thb]
\caption{Free-form natural language query categories examples with answer API and tool}
  \label{tab:q_cat_example}
\begin{tabularx}{\textwidth}{p{0.5cm}p{5cm}p{6.2cm}p{1cm}}
 \hline
$Q_c$ & Example query & Answered/ Recommended API & Tool \\ \hline
Q1 & How can I get commmunity from misp instance? & \footnotesize{pymisp.PyMISP.get\_community(community, pythonify=False)} & MISP \\\hline

Q2 & How to perform a lookup? & \footnotesize{limacharlie.Search.Search.query(iocType,iocName, 
info,isCaseInsensitive=False, isWithWildcards=False)} & Limacharlie \\\hline

Q3 & Remove intrusion detection \& response rule from organization in Limacharlie & \footnotesize{delete /rules/{oid}} & Limacharlie \\\hline
Q4 & How to listify payloads available? & \footnotesize{ limacharlie.Payloads.Payloads.list()} & Limacharlie \\\hline
Q5 & How to remove yara rule? & \footnotesize{limacharlie.Replicants.Yara.removeSource 
(sourceName)} & Limacharlie
\\ \hline
\end{tabularx}
\hspace{-10pt}
\end{table}

\begin{figure*}[thb]
  \centering
  \includegraphics[width =\textwidth]{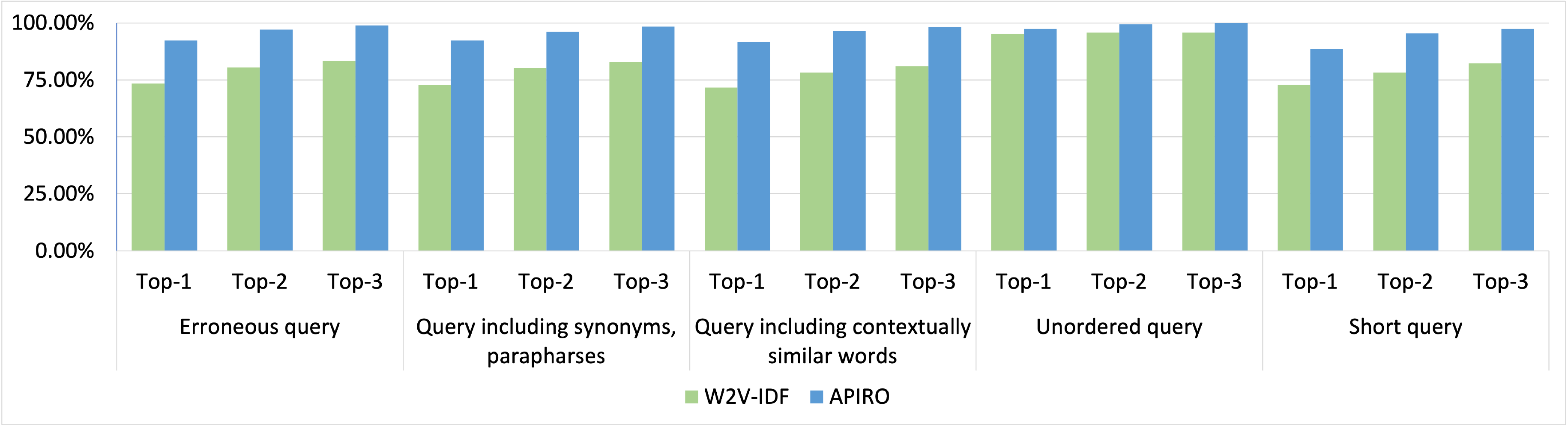}
  \caption{Performance of answering various free-form natural language query}
  \label{fig:rq4_query}
   \vspace{-10pt}
\end{figure*}

Fig. \ref{fig:rq4_query} depicts that APIRO significantly outperformed W2V-IDF in terms of Top-1 Acc, with improvement of 25.6\%, 26.8\%, and 27.9\% for erroneous query (spelling and typing mistakes) (Q1), query including synonyms, paraphrases (Q2), and query including contextually similar words (Q3), respectively. For Q1, Q2, and Q3, the Top-1 Acc of APIRO is adjacent to each other and is equal or above 91.6\%. For APIRO, among all the query categories, unordered query (Q4) category has the maximum Top-1 Acc (97.4\%). This result implies that presence of the significant terms is more important than the grammatical order of the words for the recommendation. For instance, MISP have two distinct APIs description of `disable a warninglist' and `enable a warninglist'. A query `warninglist disable' is less ambiguous as the disable term is present in the query for recommending the specific API. However, for APIRO among all the query categories,`short query' (Q5) category has the minimum Top-1 Acc (88.5\%). Following the above MISP example, a query just mentioning `warninglist' makes it ambiguous which API is required. Hence, it is more challenging to answer short queries compared to other query categories such as erroneous, unordered, or query with synonyms. 

It is to be noted that, with the increasing value of k, the recommendation accuracy of APIRO is improved for every query category. The minimum accuracy among all query categories are 88.5\%, 95.4\%, and 97.5\% for Top-1, Top-2, and Top-3, respectively, which demonstrates the increasing accuracy of APIRO with the increasing value of k. Our experimental results demonstrate that APIRO can successfully recommend API list with high accuracy for various types of free-form natural language queries. 


\section{Discussion} \label{sec: threats}
In this section, we present the strength of APIRO and the threats to validity.

\subsection{Strength of APIRO}
To support SOC teams for integration and orchestration of diverse security tools and automation of incident response activities, we propose a deep learning-based framework, APIRO, to search for the required API information. The strength of the proposed framework is summarized as follows:

\subsubsection{Data augmentation to mitigate data availability constraint}
Analysis of RQ1 shows that enriching data from different security tools using augmentation techniques is effective to mitigate the data availability constraint. This has made our framework a stand-alone solution based on API documentation as the framework does not use user-contributed data (e.g., SO, GitHub). We have synthesized the augmented data for each API using data augmentation techniques from the labeled data of API documentation (i.e., API documents include API with corresponding description, parameter, and return value representing labeled data). Hence, the augmented data mitigated the requirement of time and effort-intensive manual labeling of a huge amount of training data for the learning model. Though all the augmentation techniques are not suitable to preserve the semantics, our proposed generalized APIRO framework facilitates the selection of appropriate augmentation techniques. Besides, we have provided an immutable word corpus for security data augmentation. Moreover, our selection analysis has provided an extensive list of suitable data augmentation techniques for the security domain.

\subsubsection{Unified framework for heterogeneous data} The results of RQ2 demonstrate that APIRO has mitigated the data heterogeneity challenge by building a CNN-based model on a unified corpus of the heterogeneous security tool API data. While the existing approaches (\cite{from_word_to_doc, biker}) have built language-specific word embedding model (e.g., Java-specific model), it is not feasible for SOC to build individual tool-specific model for the large number of security tools that SOC team use. Hence, our unified framework provides them with the recommendation support tool from where they can ask queries regarding any security tool for different APIs. Besides, APIRO generates the features automatically from the unified corpus and avoides any dependency on costly hand-crafted feature generation.

\subsubsection{Semantic variation to answer natural language query} The evaluation analysis of RQ3 presents that APIRO is efficient to answer diverse free-form natural language query categories (e.g., query with synonyms or paraphrases). Since API documentation has `short description' nature, we enriched the corpus with semantic variation by applying the augmentation techniques to answer free-form queries. For example, the word `delete' was augmented with various semantic variants such as `remove', `suspend', and `erasure' in our data after applying augmentation techniques. Our framework automatically enriches API data providing words with similar functionality so that queries with different word variants can be answered. Hence, unlike the existing approach \cite{verb_match}, APIRO does not depend on manual functionality labeling and categorization to map words according to the same functionality for API recommendation.

We have proposed a deployable end-to-end framework, APIRO for the automatic recommendation of security tool APIs. We have demonstrated that our framework is fully functioning and evaluated the effectiveness of the framework using quantitative evaluation.


\subsection{Qualitative Analysis}
Evaluation result demonstrates that APIRO outperforms the state-of-the-art approach by a substantial margin in terms of accuracy. Only APIRO could recommend the correct API as top-1 result for 22\% of the queries, while W2V-IDF did not recommend the correct API for these queries. We manually checked the results of the test queries and found that both APIRO and W2V-IDF can easily provide the correct API if the query includes all the important keywords to distinguish that API from others. For example, a query including the keywords `read, single, pcap' helps both the approach to identify the correct API for reading a single pcap. Since the main keywords given in the query helps to distinguish with the other APIs such as for reading pcaps from a file or reading pcaps under a directory. 

The analysis further shows that APIRO can also recommend correct API for query types that may not have the exact keywords such as the use of different words for presenting the same concept, spelling mistakes, missing keywords, etc. These types of queries are challenging for W2V-IDF, as W2V-IDF performs word to word matching for finding similarity score and the similarity score can be comparatively low to recommend the correct API. For example, a query `How can I get commmunity from misp instance?' have spelling mistake of the term `community' to `commmunity'. W2V-IDF calculating the similarity score with the API documentation gets almost similar score for the query with multiple API descriptions such as `\textit{Get an attribute from a MISP instance'}, `\textit{Get an community from a MISP instance}', `\textit{Get an event from a MISP instance}', `\textit{Get a taxonomy from a MISP instance}', and `\textit{Get an object from the remote MISP instance}'. The keyword `community' is mainly distinguishing the correct intended API from other relevant APIs, but the corresponding query has spelling mistake in the word `commmunity'. Thus, W2V-IDF can not directly map the query to the correct API with the static similarity score calculation. However, APIRO successfully recommends the correct API. Use of subword information keeps the words `commmunity' and `community' closer in the vector space.

We hypothesize that APIRO learns to distinguish the APIs from one another, as during training APIRO learns different augmented representations of the description for each API. API documentation itself may have spelling mistake. For example, API documentation has an API description ``Performa a search", where performa is actually ``perform". A query to find that API is “How to perform a lookup?”. W2V-IDF can not directly map  the pre-processed words `performa' and `perform' as they will be considered as different word. Besides, `lookup' is not in the API description in documentation, so semantic similarity will be low to provide the correct API by W2V-IDF. However, APIRO is able to answer as it uses fastText that creates embedding of a word using its subword information. Hence, `performa' and `perform' will have similar embedding using fastText and the synonym lookup for search will be learned by the model by using data augmentation techniques which help to recommend the correct API.\\ 


Though APIRO provides the best performance compared to the state-of-the-art baseline, we found incorrect API recommendations for some queries. It is challenging for APIRO to answer queries with extreme rare terms. Given the relevant sub-word information availability in the training data, fastText can create embedding for OOV words. However, for extreme rare terms, if the relevant sub-word information is not available in the training data, fastText may fail to create the corresponding embedding. Besides, extreme rare terms  may not be produced by the applied data augmentation techniques to enrich the learning model. For example, to add a false positive rule, consider a query `How can I insert fp rule?'. With current augmentation techniques, the variation of fp is not captured or enriched. Hence, for this query APIRO provides API for adding FIM rule as top-1 result, while API for adding false positive rule as top-5 result. However, a recent data augmentation technique called `Reserved Word Augmenter' \cite{nlpaug} can be used to address this challenge. This augmenter applies target word (e.g., false positive) replacement with a list of swappable tokens (e.g., [FP, false-positive]) treated as same meaning for augmentation. Still there is a lack of any existing open-source corpus with such swappable tokens and thus require effort-intensive manual creation of such corpus for augmentation, which can be explored in the future.
APIRO and W2V-IDF both struggled with short queries having implicit context to distinguish among different APIs for recommendation. For example, a query `how to append a new rule?' is missing the context that if there is any specific type of rule to be considered. As there are different types of rules available for different security tools such as detection and response rule, FIM rule, log collection rule, and false positive rule, APRIO struggled with recommending API for queries with such implicit context.


\subsection{APIRO framework and beyond} A lack of knowledge about which tools to use and how to integrate those tools for executing IRP is one of the key reasons of inefficient (i.e., huge time delay) cyber-incident mitigation in SOC \cite{SOC_SLR}. To the best of our knowledge, little support has been provided to practitioners (SOAR developers) and users (SOC team) desiring to integrate and orchestrate the tool activities and automate incident response, thus forcing them to resort to the tedious task of manual process of retrieving API information from documentation. For this reason, APIRO is designed to stimulate API knowledge extraction by reducing time spent to dig through diverse data sources and long waiting time in Q\&A sites. SOAR developers can use APIRO to learn APIs of a security tool to integrate that tool in a SOAR platform. Besides, integration of new security tools can be required for the deployment of SOAR in a new organization's SOC environment. Moreover, SOC or SOAR team can use APIRO to define, update, or execute IRP for new incidents, new tools, or new APIs of an existing tool to cope up with the ever-evolving threat landscape. An existing survey \cite{soar_rep_2017} on the SOC team presented the top 3 challenges faced by the SOC team for security incident response are working with a large number of security tools, responding to a huge volume of incidents, and the critical time constraint for responding to those incidents. Consider handling a new zero-day attack for which there is no IRP available. For this zero-day attack, defining the IRP activities and to perform those activities the manual selection of tools and APIs can be highly challenging for the SOC team given the critical time constraint and the huge number of the available tools and the corresponding APIs. Hence, our proposed API recommender system can provide potentially valuable support in such time-critical security incident response scenarios to retrieve API data. Moreover, novice SOAR developers and SOC team members can benefit the most from the use of such recommender to retrieve API information for learning and using the APIs to perform tool integration and incident response action.\\

APIRO is trained on API description scraped from the API documentation with the corresponding API as the label for that description. To add new data sources of different formats, new adapters are required for APIRO, which include the data scraping script of the new data format. The adapter is expected to scrape new data that is the description of APIs with the corresponding API as label, parameters, and return value. Hence, manual labelling is not required to add new API data. To include new APIs of a security tool from an existing API documentation format, the existing adapter can be reused. However, an approach that is working on data sources of different formats would similarly require adding scripts to scrape the data as a basic step of data collection. For example, the existing works \cite{sec_orch_caise, islam2019ontology} that use different security tool data for different other purposes (e.g., security tool ontology construction) also require to collect data using different scrapers for different data formats. Future research can focus on the development of unified adapter for automated (e.g., pattern matching or keyword mapping approach) API data scraping from several security tools’ documentation of diverse formats. 

Besides, to add a new tool's data, it may require to select immutable words from the Noun list of the documentation that are specific to that tool. However, it mainly requires to add the tool-specific words as common security related words are usually already captured from the existing tools. Thus, the size of the Noun list to be analysed reduces drastically, which significantly reduces the required effort and time for new immutable word selection. Besides, the use of immutable or unchangeable words for a learning model is a common practice in the existing literature \cite{concept_drift, puminer}.\\

A number of research opportunities exist for extending the utility of APIRO. The existing works \cite{islam2019ontology, sec_orch_caise} manually created a semantic knowledge base of security tools capabilities for automatic integration of security tools. APIRO can be extended to automatically create the knowledge base from the API documentation. Moreover, researchers can focus on auto-generation of the API entities (e.g., parameters, input, and output) to automatically execute the selected APIs for incident response. Moreover, our proposed framework can be extended for automated question answer recommendation for other security tool artifacts (e.g., user guide, tutorials). 

There are interesting research areas that are not within the scope of this paper but can be explored to analyse their applicability in APIRO in the future. Using NLP recurrent linguistic patterns is widely used in different applications (e.g., intention mining \cite{intention_mine2}, app review classification \cite{nlp_pattern_app_review}) in the existing literature. The manual identification of recurrent linguistics patterns (e.g., [someone] should add [something]) require a large amount of time and manual effort. Since creating linguistic patterns manually requires substantial effort, the lack of generalizability may limit the applicability of this approach in our domain. Hence, the generalization of the identified patterns (i.e., if identified linguistic patterns from one security tool can be generalized to other diverse security tools) can be investigated in future work. APIRO uses char n-grams vectors to create the word embedding vector using fastText and leverage n-grams (3-gram to 5-gram) while convoluting using different filter sizes. Future research can focus on the interpretability of the CNN prediction result by identifying important n-gram sequences in the data based on the intermediate output of the neural network \cite{intention_mine1}. Furthermore, the effect of having a higher weight on the frequently-used words (e.g., top-100, top-1000) in APIRO can be investigated in the future.

\subsection{Threats to Validity}
Our evaluations were conducted on three diverse security tools that have API documentation freely available. The enrichment of the security tool API corpus with new security tool API data is an incremental and continuous process. Our framework demonstrates to consistently handle the synonyms, paraphrases, spelling/typing mistakes, contextually similar words, short, and unordered queries well even with limited data for all query categories. Still, the dataset might not fully represent every possible query and APIRO may not work for queries with extreme rare terms. However, our model is re-trainable to deal with such cases as we proposed a generalized APIRO framework, where more augmented data from the existing augmentation techniques and new augmentation techniques evolved in the literature can be easily adopted. To mitigate the threat related to the performed manual analysis of meaningful queries and the nouns of the security tools, we reported the Cohen's Kappa (0.66 and 0.75, respectively), which showed moderate agreement between the annotators. Our optimal hyper-parameters for deep learning-based model of APIRO, might not guarantee the best performance for recommending the relevant security tool APIs since it is impossible to tune the hyper-parameters of the DL model using an infinite number of available configurations. For minimizing this threat, we followed the state-of-the-art approach to choose optimal hyper-parameters and then ran our model multiple times with 10-fold cross-validation. Besides, we chose CNN-based model for security tool API recommendation as it achieved optimum results in the existing research \cite{cnn_sentence, tosem_ans}. Our evaluation does not give insight in the performance of other deep learning models (e.g., BERT \cite{bert}, RCNN \cite{rcnn}, BLSTM-2DCNN \cite{BLSTM-2DCNN}) as our focus in this paper is to provide a foundation framework for automated security tool API recommendation. Future studies can be done on applying and comparing other deep learning models for recommending security tools API.
\section{Conclusion} \label{sec: concl}
This paper presents a novel learning based framework, APIRO, to assist SOC team in finding suitable security tools' APIs, considering a SOAR platform environment in a SOC. The proposed approach allows SOC team to select suitable APIs that are required to integrate security tools and execute IRPs. We have introduced a data collector to retrieve security tools API information from a wide variety of documentations and a data augmenter to enrich the retrieved API task descriptions. A Convolutional Neural Network (CNN)-based learning model is proposed to learn the data heterogeneity of security tools with the assistance of a word embedding model. The word embedding model has been built using the augmented security tools' API data. The proposed CNN model successfully recommends the top-3 relevant APIs for different tasks in response to natural language queries.

We have demonstrated the effectiveness of APIRO by implementing a Proof-of-Concept (PoC) system. The PoC system has gathered 815 unique APIs from three security tools, that are enriched to 24,420 API descriptions by applying 36 data augmentation techniques.Our analysis of the results shows that using word-embedding and CNN to learn semantic variation in security tool API description is 26.9\% more efficient than using W2V-IDF similarity score-based approach. Further, the effectiveness of APIRO that is the minimum accuracy to recommend API is  88.5\%, 95.4\%, and 97.5\% for top-1, top-2, and top-3, respectively in response to different natural language query categories. We assert that our proposed approach can minimize the overhead of SOCs involved in manually finding APIs and effectively recommend APIs of different security tools.

In the future, our aim is to build an automated recommendation system (e.g., web interface) to support the SOC team and SOAR developers to use APIRO for searching APIs while using heterogeneous security tools. It will support incremental training on new data and allow user to incorporate data from more security tools (e.g., Splunk, OSSEC, Cuckoo). Besides, we intend to start a research project for developing and applying a framework to empirically evaluate APIRO and similar frameworks/tools. We also intend to release APIRO as open source platform so that other researchers and SOC teams can also use and evaluate the tool and provide feedback. Moreover, researchers can extend APIRO to automate the generation of API entities for auto integration of security tools in a SOAR platform. Future research can also focus on using APIRO recommended APIs to automate execution of an incident response process when new activities with new security tools are integrated.

\begin{acks}
The work has been supported by the Cyber Security Research Centre Limited whose activities are partially funded by the Australian Government’s Cooperative Research Centres Programme.
This work has also been supported with super-computing resources provided by the Phoenix HPC service at the University of Adelaide. 
\end{acks}

\bibliographystyle{ACM-Reference-Format}
\bibliography{sample-base}

\appendix
\section{Appendix}
This section presents the identified immutable word corpus from our analysis and the details of the data augmentation techniques that are used in our experiment.
\subsection{Immutable word corpus}
Identified immutable word corpus from our analysis:{\footnotesize cloud, ransomware, eth0, afpacket, checksum, read-file, snort, pid, pcaps, tcpdump, all\_orgs, console, -q, -d, -u, pptp, lro, fim, package, mitre, limacharlie, bpf, -l, chrome, datagram, mpls, blob, payload, malware, megabyte, verbose, stub, linux, ip, webhooks, pcap, vlan, galaxy, attribute\_identifier, icmp, erspan, api, clone, macos, ascii, memcap, snaplen, pull, user\_id, sha1, apis, pcre, exception, sniffer, misp\_entity, -e, yara, ioc, -x, misp-modules, iptables, iocs, param, -g, firewall, socket, to\_ids, -i, thresholding, -t, bytesio, header, -v, mac, pymisp, webhook, -m, io, mispevent, int, dlls, import\_server, encapsulation, multicast, hids, initd, md5, daq, misp, attribute, protocol, tag, ipv4, key, version, pythonified, default, decoder, -snaplen, decodes, library, audit, function, hash, asn1, rpz, chroots, -c, -y, http, gre, dns, exfil, str, metadata, cached, cronjobs, configs, nfs, dylibs, tcp, xml, hostnames, sensor, buffer, uuid, replay, -n, pseudofile, offload, warninglist\_id, ttl, byte, syslog, sha256, bit, json, suricata, hostname, get\_sync\_config, port, cmg, -p, spotcheck, fork, zmq, github, object, launchctl, teredo, nids, csv, encoder, cache, python, -o, warninglist, ipv6, log, stdout, object\_relation, -f, sharing\_group.}

\subsection{Data augmentation techniques details}
Table \ref{tab:append_dat_detail} presents the summary of the data augmentation techniques with corresponding augmentation approaches, actions, data source details, and the existing publicly available data source links that we have used for our experiment. In this table, Ins: Insert, Subs: Substitute, SWord: Subword, CCrawl: Common Crawl, and Gword: Gigaword.

{\footnotesize
\begin{longtable}{|p{1cm}|p{1.4cm}| p{4.6cm}| p{3cm}|p{2cm}|p{1cm}|} \caption{Summary of the data augmentation techniques with corresponding augmentation approaches, actions, data source details, and available resource links} \label{tab:append_dat_detail} \\\hline
DAT & DAT name & Data Source Details & Data Source Link & Aug Approach & Action \\ \hline
 \endfirsthead
\multicolumn{6}{@{}l}{\ldots continued}\\\hline
DAT & DAT name & Data Source Details & Data Source Link & Aug Approach & Action\\\hline
\endhead 
\hline
\multicolumn{6}{r@{}}{continued \ldots}\\
\endfoot
\endlastfoot
DAT 1 & Swap Word & - & - & \multirow{2}{1.8cm}{$Aug_1$. RandomWordAug} & swap \\ \cline{1-4} \cline{6-6} 
DAT 2 & Delete Word & - & - &  & delete \\ \hline

DAT 3 & Spelling Augmenter & Pre-defined spelling mistake dictionary. & \url{https://github.com/makcedward/nlpaug/blob/master/nlpaug/res/word/spelling/spelling\_en.txt} & $Aug_2$. SpellingAug & substitute \\ \hline

DAT 4 & Split Words & - & - & $Aug_3$. SplitAug & split \\ \hline

DAT 5 & Synonym Wordnet Subs & Wordnet is a vast lexical English database that includes the cognitive synonym sets of words. & \url{https://wordnet.princeton.edu/} & {$Aug_4$. SynonymAug} & \multirow{2}{*}{substitute} \\ \cline{1-4}
DAT 6 & Synonym PPDB Subs & PPDB database contains millions paraphrases to improve language processing by making systems more robust to language variability and unseen words. Six sizes, from S up to XXXL are available. We used XXXL which includes the maximum paraphrases. & \url{http://paraphrase.org/\#/download} &  &  \\ \hline \pagebreak
DAT 7 & Tf\_Idf Ins & \multirow{2}{4.6cm}{We trained TF-IDF model on our API corpus to augment data using TF-IDF statistics.} & \multirow{2}{3cm}{\url{https://github.com/makcedward/nlpaug/blob/master/example/tfidf-train\_model.ipynb}} & {$Aug_5$. TfIdfAug} & insert, substitute \\ \cline{1-2} 
DAT 8 & \specialcell{Tf\_Idf \\Subs} &  &  &  &  \\ \hline

DAT 9 & Word2vec Googlenews Ins & \multirow{2}{4.6cm}{Word2vec Googlenews includes 3 million word vectors which are pre-trained using portion of Googlenews corpus.} & \multirow{2}{3cm}{\url{https://code.google.com/archive/p/word2vec/}} & \multirow{2}{2cm}{$Aug_6$. WordEmbsAug} & \multirow{2}{1cm}{insert, substitute} \\ \cline{1-2}
DAT 10 & Word2vec Googlenews Subs &  &  &  &  \\ \cline{1-4}
DAT 11 & Fasttext Wikinews Ins & \multirow{2}{4.6cm}{Fasttext Wikinews includes 1 million word vectors pre-trained using Wikipedia (2017), UMBC webbase, and statmt.org news corpus.} & \multirow{2}{3cm}{\url{ https://dl.fbaipublicfiles.com/fasttext/vectors-english/wiki-news-300d-1M.vec.zip}} &  &  \\ \cline{1-2}
DAT 12 & Fasttext Wikinews Subs &  &  &  &  \\ \cline{1-4}
DAT 13 & Fasttext Wiki-News SWord Ins & \multirow{2}{4.6cm}{Fasttext Wikinews SWord includes 1 million word vectors pre-trained including subword information using Wikipedia (2017), UMBC webbase, and statmt.org news corpus.} & \multirow{2}{3cm}{\url{https://dl.fbaipublicfiles.com/fasttext/vectors-english/wiki-news-300d-1M-subword.vec.zip}} &  &  \\ \cline{1-2}
DAT 14 & Fasttext Wiki-News SWord Subs &  &  &  &  \\ \cline{1-4}
DAT 15 & Fasttext CCrawl Ins & \multirow{2}{4.6cm}{Fasttext CCrawl includes 2 million word vectors pre-trained using Common Crawl corpus.} & \multirow{2}{3cm}{\url{https://dl.fbaipublicfiles.com/fasttext/vectors-english/crawl-300d-2M.vec.zip}} &  &  \\ \cline{1-2}
DAT 16 & Fasttext CCrawl Subs &  &  &  &  \\ \cline{1-4}
DAT 17 & Fasttext CCrawl SWord Ins & \multirow{2}{4.6cm}{Fasttext CCrawl SWord includes 2 million word vectors pre-trained including subword information using Common Crawl corpus.} & \multirow{2}{3cm}{\url{https://dl.fbaipublicfiles.com/fasttext/vectors-english/crawl-300d-2M-subword.zip}} & \multirow{2}{2cm}{$Aug_6$. WordEmbsAug} & \multirow{2}{1cm}{insert, substitute} \\ \cline{1-2}
DAT 18 & Fasttext CCrawl SWord Subs &  &  &  &  \\ \cline{1-4}
DAT 19 & Fasttext CCrawl \& Wiki Ins & \multirow{2}{4.6cm}{Fasttext CCrawl \& Wiki is trained using Common Crawl and Wikipedia corpus using fastText approach.} & \multirow{2}{3cm}{\url{https://fasttext.cc/docs/en/crawl-vectors.html}} &  &  \\ \cline{1-2}
DAT 20 & Fasttext C. Crawl \& Wiki Subs &  &  &  &  \\ \cline{1-4}
DAT 21 & Glove Wiki+Gword 50d Ins & \multirow{4}{4.6cm}{Glove Wiki+Gword includes 400K vocabulary trained using Wikipedia (2014) and Gigaword 5 corpus with vector dimensions of 50d, 100d, 200d, \& 300d. We used both 50d and 300d, the minimum and maximum dimension size to check the variation.} &
\multirow{2}{3cm}{\url{http://nlp.stanford.edu/data/glove.6B.zip}} &  &  \\ \cline{1-2}
DAT 22 & Glove Wiki+Gword 50d Subs &  &  &  &  \\ \cline{1-2}
DAT 23 & Glove Wiki+Gword 300d Ins &  &  &  &  \\ \cline{1-2}
DAT 24 & Glove Wiki+Gword 300d Subs &  &  &  &  \\ \cline{1-4}\pagebreak
DAT 25 & Glove C. Crawl 300d Ins & \multirow{2}{4.6cm}{Glove CCrawl includes 1.9M vocabulary trained using Common Crawl corpus} & \multirow{2}{3cm}{\url{http://nlp.stanford.edu/data/glove.42B.300d.zip}} &  &  \\ \cline{1-2}
DAT 26 & Glove C. Crawl 300d Subs &  &  &  &  \\ \cline{1-4}
DAT 27 & Glove Twitter 200d Ins & \multirow{2}{4.6cm}{Glove Twitter includes 1.2M vocabulary trained using Twitter corpus with vector dimensions of 25d, 50d, 100d, \& 200d. We used 200d dimension vectors as it contains maximum dataset size.} &
\multirow{2}{3cm}{\url{http://nlp.stanford.edu/data/glove.twitter.27B.zip}} & \multirow{2}{2cm}{$Aug_6$. WordEmbsAug} & \multirow{2}{1cm}{insert, substitute}  \\ \cline{1-2}
DAT 28 & \specialcell{Glove \\Twitter 200d\\ Subs} &  &  &  &  \\ \hline 
DAT 29 & Bert Base Ins & \multirow{2}{4.6cm}{Bert Base trained on lower-cased English text using 12 layers and 110M parameters.} & \multirow{2}{3cm}{\url{https://huggingface.co/transformers/pretrained\_models.html}} & {$Aug_7$. ContextualWordEmbsAug} & \multirow{2}{1.2cm}{insert, substitute} \\ \cline{1-2}
DAT 30 & Bert Base Subs &  &  &  &  \\ \cline{1-4} 
DAT 31 & Distilbert Ins & \multirow{2}{4.6cm}{Distilbert distilled from the bert-base-uncased model uses 6 layers and 66M parameters.} & \multirow{2}{3cm}{\url{https://github.com/huggingface/transformers/tree/master/examples/distillation}} &  &  \\ \cline{1-2}
DAT 32 & \specialcell{Distil-\\bert\\ Subs\\} &  &  &  &  \\ \cline{1-4}
DAT 33 & Roberta Base Ins & \multirow{2}{4.6cm}{Roberta Base uses BERT-base architecture and have 12 layers and 125M parameters.} & \multirow{2}{3cm}{\url{https://github.com/pytorch/fairseq/tree/master/examples/roberta}} &  &  \\ \cline{1-2}
DAT 34 & Roberta Base Subs &  &  &  &  \\ \cline{1-4}
DAT 35 & Distilroberta-Base Ins &\multirow{2}{4.6cm}{Distilroberta-Base distilled from roberta-basecheckpoint and uses 6 layers and 82M parameters.} & \multirow{2}{3cm}{\url{https://github.com/huggingface/transformers/tree/master/examples/distillation}} &  &  \\ \cline{1-2}
DAT 36 & Distilroberta-Base Subs &  &  &  &  \\ \hline
\end{longtable}
}

\end{document}